\documentclass[a4paper,12pt]{article}

\usepackage{geometry}
\usepackage{amsmath,amssymb,amsfonts}
\usepackage{graphicx}
\usepackage{hyperref}
\usepackage{cite}
\usepackage{enumitem}
\usepackage{caption}
\usepackage{float}
\usepackage{booktabs}
\usepackage{adjustbox}
\usepackage{authblk}
\usepackage[inkscapelatex=false]{svg}
\usepackage{array}
\usepackage{makecell}
\usepackage{longtable}
\usepackage{multirow}
\usepackage{array,longtable}

\title{Optimal Strategy in the Werewolf Game: A Theoretical Study}
\date{}
\author{Shitong Wang\thanks{\href{shitwang@mail.uni-mannheim.de}{shitwang@mail.uni-mannheim.de}}}
\affil[]{Univeristy of Mannheim}

\begin{document}
\maketitle  
\begin{abstract}
In this paper, we investigate the optimal strategies in the Werewolf Game—a widely played strategic social deduction game involving two opposing factions—from a game-theoretic perspective. We consider two scenarios: the game without a prophet and the game with a prophet. In the scenario without a prophet, we propose an enhanced strategy called ``random strategy+'' that significantly improves the werewolf group’s winning probability over conventional random strategies. In the scenario with a prophet, we reformulate the game as an extensive-form Bayesian game under a specific constraint, and derive the prophet’s optimal strategy that induces a Perfect Bayesian Equilibrium (PBE). This study provides a rigorous analytical framework for modeling the Werewolf Game and offers broader insights into strategic decision-making under asymmetric and incomplete information.
\end{abstract}

\newpage
\section{Introduction}
\subsection{Game Description}
\textit{Werewolf Game}, also known as \textit{Mafia Game}, originated from a Russian social deduction game created by \textit{Dimitry Davidoff} in 1986. It has since evolved into a widely played game of strategic reasoning, deception, and incomplete information. Typically involving 8–16 players, each is assigned a hidden role belonging to one of two opposing factions: the \textit{citizen group} and the \textit{werewolf group}. 

The citizen group aims to eliminate all werewolves via public daytime voting, while the werewolf group strives to reduce the number of citizens to zero by secretly eliminating one player each night. To add complexity and entertainment value to the game, the citizen group is often assigned a prophet who can check one player's group affiliation each night.

This game dynamic creates a rich environment for studying strategic decision-making under uncertainty. Players must reason not only with their own information, but also infer the intentions and beliefs of others, balancing short-term risks with long-term outcomes. In particular, the timing of the prophet’s information revelation plays a pivotal role: revealing too early risks premature death and wasted influence, while revealing too late may allow the werewolf group to seize control.

\subsection{Literature Review}
Academic research on the Werewolf Game typically falls into three main categories:
\begin{enumerate}
    \item \textbf{Probability and Game Theory:} Studying relevant strategies, equilibria, and winning rates of both groups under specific game settings.
    \item \textbf{Social and Behavioral Sciences:} Studying phenomena in social psychology within games, such as group behavior, persuasion, and deception. This includes analyzing how players interact during the game, and examining the factors that influence their decisions and actions.
    \item \textbf{Artificial Intelligence and Algorithms:} Using the Werewolf Game as a testbed for algorithm design, or leveraging in-game dialogue to train and evaluate artificial intelligence in assessing information authenticity.
\\
\end{enumerate}

This study models the Werewolf Game assuming fully rational agents and does not incorporate psychological or behavioral considerations. The subsequent literature review therefore concentrates on prior work in probability and game theory, which are most relevant to the present analysis.

\textit{Braverman} et al. (2008) suggested that in the Werewolf Game without a prophet, if both the citizen group and the werewolf group adopt random strategy, then the two groups have comparable winning probabilities when the werewolves' size is of the order of the square root of the total players' size. In the Werewolf Game with prophets, the two groups have comparable winning probabilities when the werewolves' size and total players’ size are linearly related. Regarding the game without a prophet, \textit{Yao} (2008) corrected the mathematical flaws of the probability boundary function in the original theorem of \textit{Braverman} et al. (2008) and gave a more precise recursive formula and probability upper and lower bounds of players' winning probability.    \textit{Migdał} (2013) calculated the analytical solution of winning probability for the Werewolf Game without a prophet. The above papers (\textit{Braverman} et al. (2008); \textit{Yao} (2008); \textit{Migdał} (2013)) all assumed that in the Werewolf Game without a prophet, ``random strategy'', i.e., the werewolf group eliminates a citizen randomly during the night and all players vote to eliminate a player randomly during the day is the optimal strategy for both groups, but we will revise this assumption. \textit{Bi} et al. (2016) calculated the Nash equilibrium of the game under certain limitations and conclude that the ``stealth werewolf'' strategy, i.e., werewolves pretending to be citizens, is not a good strategy. \textit{Xiong} et al. (2017) used the \textit{Game Refinement Measure}, a measure to qualify the sophistication of a game, to measure the Werewolf Game and concluded that too many players in a single game may make the game overly complex and less engaging.

This paper investigates the optimal strategies for both groups in the Werewolf Game under settings with and without the presence of a prophet.
For games without a prophet, our main contribution is the proposal of an improved strategy for the werewolf group, which achieves a higher winning probability than conventional approaches commonly used in practice.
For games with a prophet, we develop a recursive Bayesian game model that captures the prophet’s information structure and strategic considerations. Based on this model, the prophet can compute an optimal strategy under any given circumstance, including scenarios that rarely occur in standard gameplay.
 
\section{Background of the Werewolf Game}
In this section, we provide a formal introduction to fundamental rules of the Werewolf Game, which lays the foundation for our subsequent modeling and analysis.

\subsection{Role Set Configuration and Role Assignment}

At the outset of the game, the total number of players and the composition of role set (i.e., the number of players assigned to each role type) are common knowledge, while each player's specific role remains private information.  

Players are typically divided into two opposing factions: the citizen group and the werewolf group. Within the citizen group, those without special power are referred to as \textit{villagers}, while others may possess special powers. For example, a citizen who can check another player’s group affiliation (citizen or werewolf) once per night is referred to as a \textit{prophet}. Some versions of the game also permit werewolves to possess special powers, but such extensions are not considered in our model. A simple role set might be two villagers, one prophet, and two werewolves, forming a five-player game.

In some variants of the Werewolf Game, the exact role set may be uncertain. Nevertheless, even in these cases, the probability distribution over possible role sets is still common knowledge.

To facilitate subsequent analysis, players are assigned serial numbers in a clockwise order. When a player is eliminated from the game, the serial numbers of all remaining players with higher numbers are adjusted downward to close the gap. For example, if player 4 is eliminated, then player 5 becomes player 4, player 6 becomes player 5, and so on.

\subsection{Gameplay Process}

With the initial setup complete, players formally begin the game according to the established rules. Although there are many versions of the Werewolf Game, each with slightly different rule sets, we present a representative version that serves as the basis for our analysis. The game proceeds in a sequence of rounds, each consisting of a night phase followed by a day phase. 

Before detailing the game procedure, we briefly introduce the possible roles in the version of the Werewolf Game considered in this paper.

\begin{table}[H]
    \centering
    \caption{Different types of roles}
    \begin{tabular}{|c|c|p{10cm}|}
    \hline
    \textbf{Role} & \textbf{Group} & \textbf{Special power} \\ \hline
    Villager & Citizen & No special power \\ \hline
    Werewolf & Werewolf & No special power \\ \hline
    Prophet & Citizen & Check one player's group affiliation every night \\ \hline
    \end{tabular}
\end{table}

On the first night, the werewolves learn each other's identities. During the night phase, the werewolf group collectively selects one player to eliminate. This player is removed from the game at the beginning of the following day. If a prophet is present, they may check the group affiliation of one player each night. It is worth noting that the werewolves’ elimination takes effect after the night; therefore, the prophet may end up checking the group affiliation of a player who is simultaneously being eliminated during that same night.

During the day phase, each player has one opportunity to publicly announce a message to all other players. These announcements occur simultaneously, and no private communication is permitted. Following the announcements, all players simultaneously vote to eliminate one player. The player receiving the highest number of votes is immediately removed from the game. In the event of a tie, one of the tied players is randomly chosen for elimination. The next night phase then begins.

Importantly, the group affiliations of eliminated players—whether eliminated by the werewolves or by public vote—are not revealed to the remaining players. This feature marks a key distinction between our game model and those studied in \textit{Braverman} et al. (2008) and \textit{Yao} (2008). In their model, the prophet voluntarily sacrifices themselves in a specific round to validate their identity and securely convey checking information to the verified citizens, assuming the availability of private communication channels. Clearly, such a strategy is not feasible under our assumptions.

The game begins with a night phase and alternates between night and day until all players from one faction—either the citizen group or the werewolf group—are eliminated. The surviving group is then declared the winner.

\begin{figure}[H]
    \centering
    \includegraphics[width=0.8\textwidth]{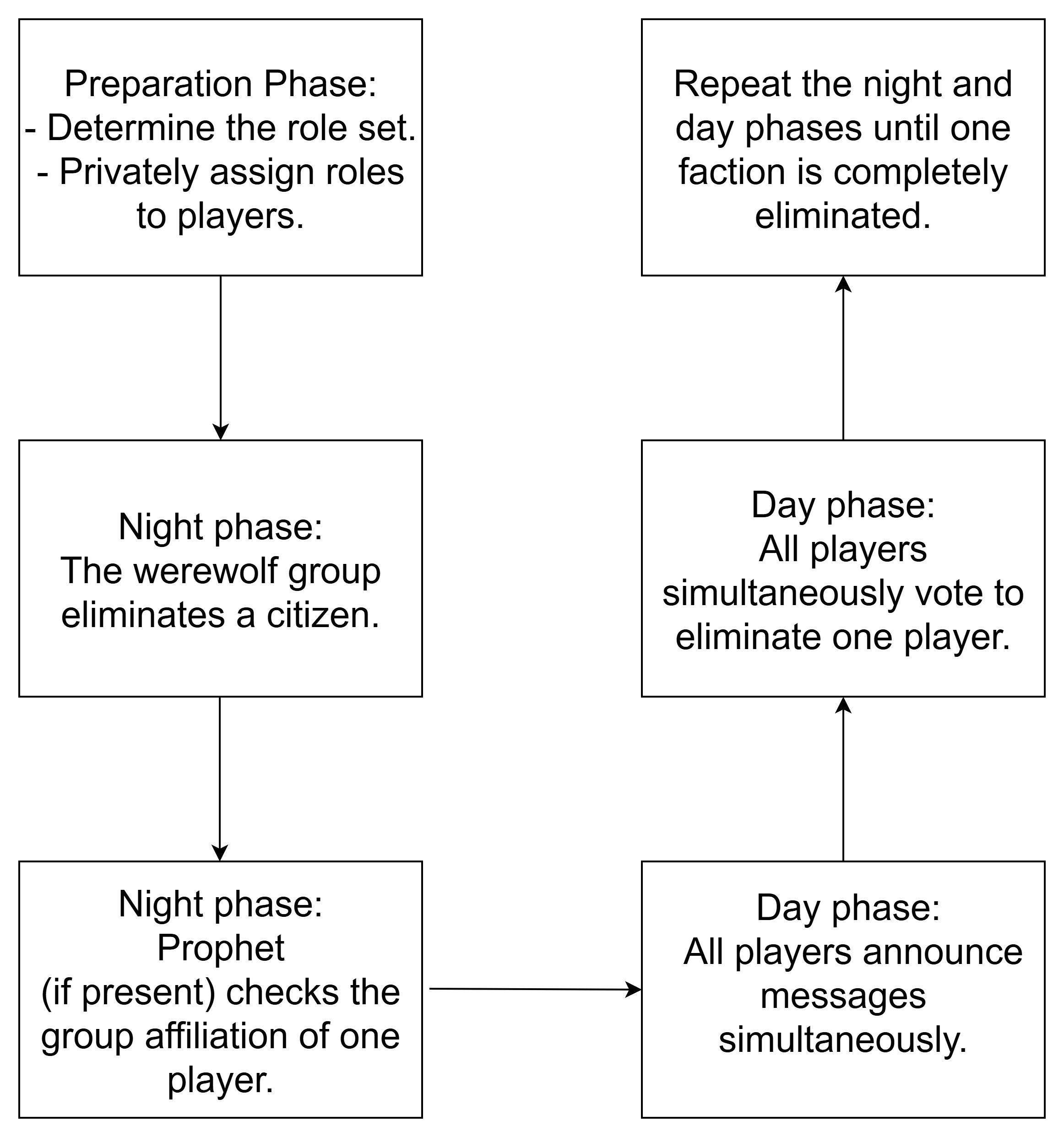}
    \caption{Process of the Werewolf Game}
\end{figure}

\section{Game without a prophet} 
 
We now turn to the simplest and most widely discussed case: the Werewolf Game without a prophet. In this setting, we assume that all citizens are villagers and that all werewolves possess no special powers.

As mentioned above, in game without a prophet, many previous studies (eg., \textit{Braverman} et al. (2008); \textit{Yao} (2008); \textit{Migdał} (2013)) have assumed that the optimal strategy for citizens (comprising only villagers) is to request all players to vote randomly during the day, while the optimal strategy for werewolves is to randomly eliminate a citizen during the night. This assumption stems from the absence of any information that would justify targeting specific players.

However, the assumption of ``all players voting randomly'' during the day encounters a practical challenge in real gameplay: it is difficult to verify whether players are truly voting at random. Werewolves, in particular, have incentives to avoid voting for their teammates and may collude to target specific players during the night. This behavior increases the likelihood of eliminating citizens while reducing the risk of werewolves being voted out. Nonetheless, werewolves can still claim that their votes were cast randomly—just like honest citizens—making it hard to distinguish between the two.

Fortunately, the citizen group has a natural mechanism to counter this issue. An implicit rule, often accepted by all players without the need for explicit agreement, is employed to enforce random voting. According to this rule, during the phase in which each player ``simultaneously announces messages'', every player simultaneously selects a natural number. The sum of all selected numbers is then computed, and the result is taken modulo the total number of players. The player whose serial number corresponds to the result of the modulo operation is designated as the target for elimination in the subsequent vote. Any player who fails to support their declared number or votes against the modulo result is immediately identified as a werewolf and eliminated during the voting in the next round.

This mechanism ensures that even though the citizens (an uninformed majority) lack knowledge about others' roles, they can still enforce fair and verifiable random voting. Meanwhile, the werewolves (an informed minority) are forced to conform to the rule to avoid suspicion and exposure.

More precisely, the process proceeds as follows:  
Each player \( i \in \{1, 2, \ldots, n\} \) simultaneously selects a natural number \( x_i \in \mathbb{N} \) during the message announcement phase. All chosen numbers are then publicly disclosed, and their sum \( S = \sum_{i=1}^n x_i \) is computed. The player whose serial number equals \( S \bmod n \) (interpreting 0 as \( n \)) is selected as the elimination target.

Formally, define
\begin{equation}
t = 
\begin{cases}
S \bmod n, & \text{if } S \bmod n \ne 0, \\
n, & \text{if } S \bmod n = 0,
\end{cases}
\end{equation}
then player \( t \) becomes the designated elimination target.
 
The strategy described above, commonly referred to as the ``random strategy'', has been regarded as optimal for both factions in prior works by \textit{Braverman} et al.\ (2008), \textit{Yao} (2008), and \textit{Migdał} (2013). However, we propose an enhanced strategy for the werewolf group that improves their winning probability, particularly in games with a small number of players.

The improved strategy operates as follows: when the number of werewolves equals the number of remaining citizens during the voting phase, if the player designated for elimination via the modulo operation is a citizen, the werewolf group simply follows the voting rule and secures an immediate victory, as their number then exceeds that of the citizens. In contrast, if the designated player is a werewolf, the werewolf group may adopt an ``all-in strategy''. This entails pre-coordinating during the previous night to unanimously vote for a specific citizen, thereby forcing a tie in the voting outcome. If the tie is resolved in favor of eliminating the targeted citizen (e.g., with probability $\frac{1}{2}$) the werewolf group wins. If the werewolf is eliminated instead, the remaining werewolves eliminate a citizen during the following night and repeat the ``all-in strategy'' in the next round, continuing this cycle until the game concludes.

Naturally, once this strategy is employed, the citizens can infer the identities of all werewolves. In the following rounds, they can coordinate directly to vote for one of the werewolves, resulting in a tie with the werewolves' votes.

We refer to the combination of the ``random strategy'' and the ``all-in strategy'' as the ``random strategy+''. We will now demonstrate that ``random strategy+'' induces a \textit{Perfect Bayesian Equilibrium} (PBE) in game without a prophet.

\begin{figure}[H]
    \centering
    \includegraphics[width=0.7\textwidth]{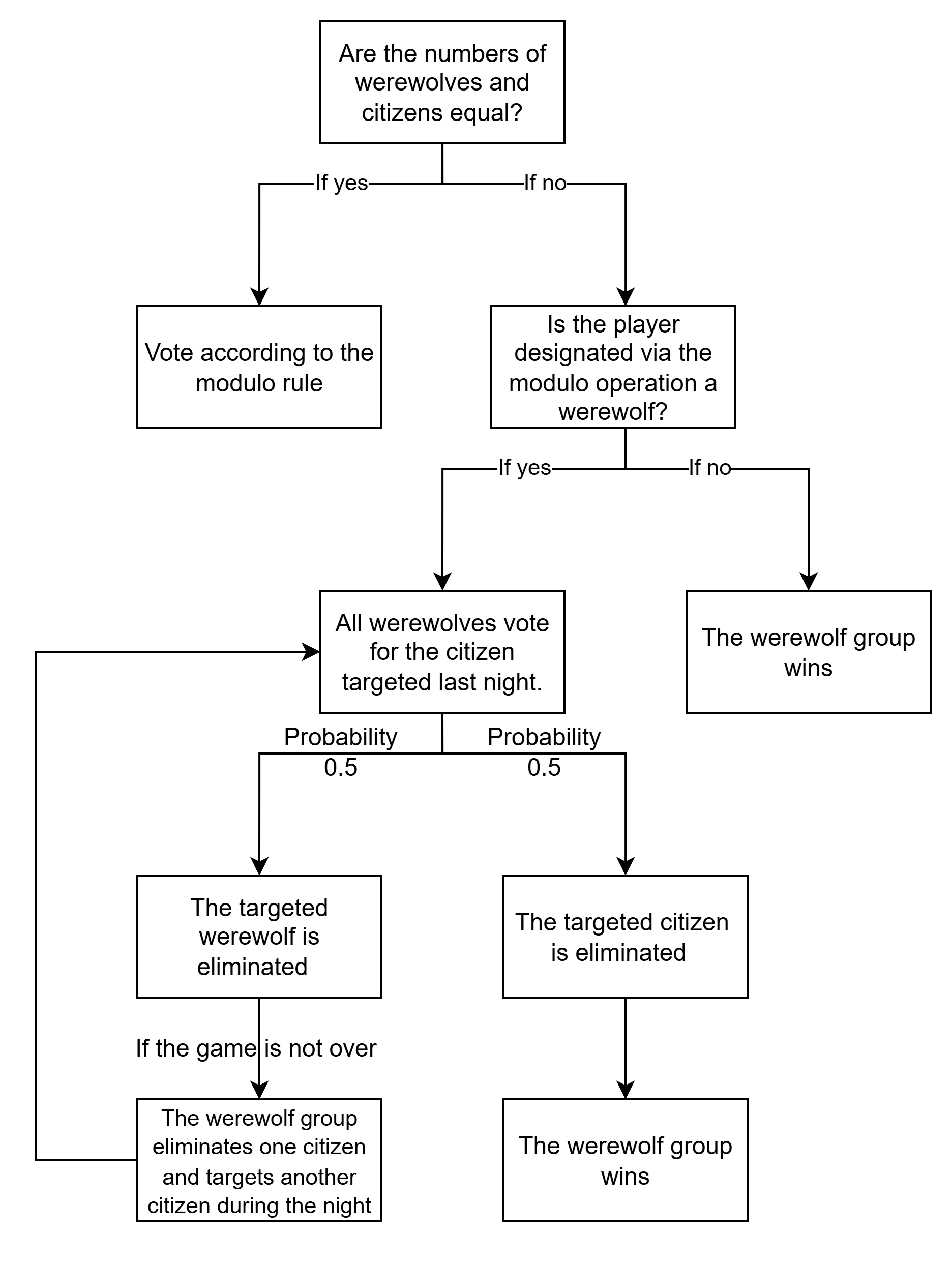}
    \caption{Schematic diagram illustrating the werewolf group's ``all-in strategy''}
\end{figure}

As discussed earlier, during the day phase, the citizen group lacks sufficient information to make informed decisions. Consequently, their only viable strategy is to ensure the integrity of the random voting process, enforced via the modulus mechanism. This makes the strategy the uniquely optimal choice for the citizen group.

We now demonstrate that ``random strategy+'' is also optimal for the werewolf group. Suppose that \textbf{at the end of the day phase and the beginning of the night phase}, there are \( n \) players remaining, of whom \( m \) are werewolves. Let \( w(n, m) \) denote the probability that the werewolf group ultimately wins the game when employing the ``random strategy+''. Then \( w(n, m) \) satisfies the following recursive formula:

\begin{align}
w(n, m) =
\begin{cases}
0, & \text{if } m = 0, \\
1, & \text{if } m \geq n - m, \\
1 - \left(\frac{1}{2}\right)^{m+1} & \text{if } n - 1 = 2m, \text{ and } n \geq 5, \\
\frac{n - 1 - m}{n - 1} w(n - 2, m) + \frac{m}{n - 1} w(n - 2, m - 1), & \text{otherwise.}
\end{cases}
\end{align}

Interestingly, when the number of werewolves is equal to the number of citizens and the designated player is a werewolf during the voting phase, for the werewolf group, employing the ``all-in strategy'' immediately is mathematically equivalent to obeying the ``random strategy'' rule until the number of werewolves reduces to 2.
From the perspective of obeying the ``random strategy'' (assuming $m>2$), we observe
 
\begin{align}
w(2m+1,m) &= \frac{2m+1- 1 - m}{2m} w(2m-1, m) + \frac{m}{2m+1 - 1} w(2m - 1, m - 1) \notag \\
&= \frac{1}{2} + \frac{1}{2} w(2m-1,m-1).
\end{align}
From the perspective of the werewolf group employing ``all-in strategy'' (assuming $m>2$), we have
\begin{equation}
w(2m+1,m)=1-(\frac{1}{2})^m=\frac{1}{2}+\frac{1}{2}(1-(\frac{1}{2})^{m-1})=\frac{1}{2}+\frac{1}{2}w(2m-1,m-1).  
\end{equation}
Therefore, in order to simplify the formula, $w(n,m)$ can also be written as:
\begin{equation}
 w(n, m) =
\begin{cases}
 0, & \text{if } m = 0, \\
\frac{7}{8}, & \text{if } n =5\text{ and }m = 2, \\
1, & \text{if } m \geq n - m, \\
 \frac{n - 1 - m}{n - 1} w(n - 2, m) + \frac{m}{n - 1} w(n - 2, m - 1), & \text{otherwise.}
\end{cases}
\end{equation}

Analogously, we can derive the recursion of the werewolf group winning probability when the werewolf group employs the ``random strategy''. Suppose that \textbf{at the end of the day phase and the beginning of the night phase} there are \( n \) players remaining, of whom \( m \) are werewolves. Let \( v(n, m) \) denote the probability that the werewolf group ultimately wins the game when employing the ``random strategy''. Then \( v(n, m) \) satisfies the following recursive formula:
\begin{equation}
v(n, m) = 
\begin{cases} 
0 & \text{if } m = 0 \\
1 & \text{if } m \geq n - m \\
\frac{n-1-m}{n-1} v(n - 2, m) + \frac{m}{n-1} v(n - 2, m - 1) & \text{otherwise}
\end{cases} 
\end{equation}
Now we prove that the ``random strategy +'' weakly dominates the ``random strategy'':\\
When \(n\) is even, the ``all-in strategy'' would never happen, then \(w(n,m)=v(n,m)\) for all \(n,m\).
When \(n\) is odd, for all \(n\geq5\) and \(m\geq2\), \(w(n,m)\) or \(v(n,m)\) can be written as the linear form of \(w(5,2)\) or \(v(5,2)\):
\begin{equation}
w(n_i,m_j)=c_{ij}+\alpha_{ij}\cdot w(5,2),
\end{equation}
\begin{equation}
v(n_i,m_j)=c_{ij}+\alpha_{ij}\cdot v(5,2).
\end{equation}
Since we get $w(5,2)=\frac{7}{8}\geq v(5,2)=\frac{3}{4}$, then $w(n,m)\geq v(n,m)$ for all odd \(n\geq5\) and \(m\geq2\).
Easy to verify, when \(n<5\) or \(m<2\), $w(n,m)=v(n,m)$.

In summary, \( w(n, m) \geq v(n, m) \) for all \( n \) and \( m \). We have proven that the ``random strategy+'' weakly dominates the ``random strategy'' in all cases for the werewolf group. The ``random strategy+'' is the optimal strategy for both groups and thus can indeed induce a PBE. The following figure illustrates the difference in the werewolf group's winning probability under these two strategies. Bar colors are used solely for visual clarity.\\
 
\begin{figure}[H]
    \centering
    \includegraphics[width=\textwidth]{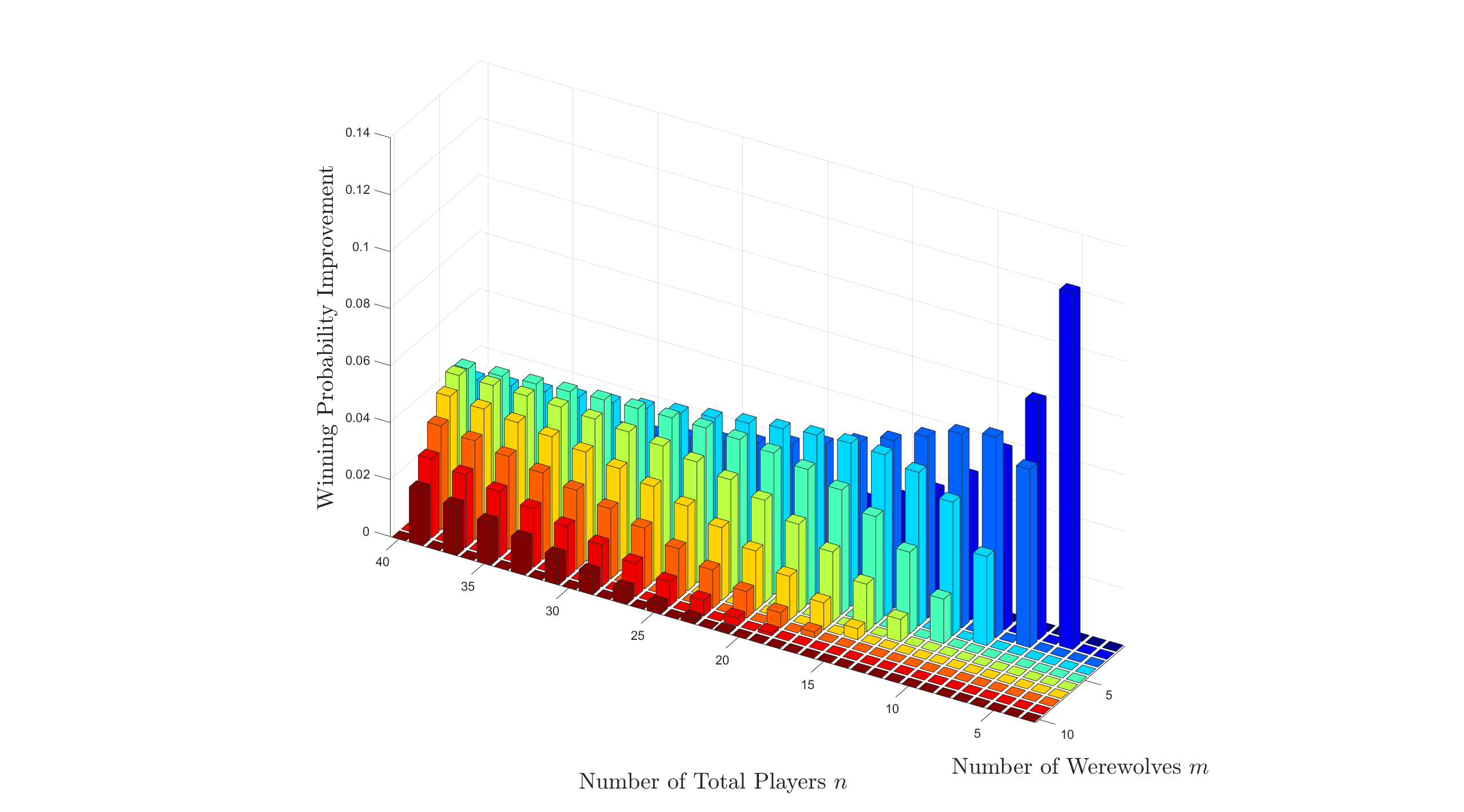}
    \caption{Winning probability improvement for the werewolf group under the ``random strategy+'' compared to the ``random strategy''}

\end{figure}
 
Finally, we prove that the werewolf group’s strategy of self-killing during the night is strictly dominated. This proof is necessary because, in all previous analyses, we implicitly assumed that the werewolves would not adopt such a strategy. Providing a formal justification for this assumption is essential to completing the proof that the ``random strategy+'' is indeed the optimal strategy for the werewolf group.

\vspace{0.5em}
\noindent\textbf{Claim 1.} In a game without a prophet, the werewolf group’s strategy of killing themselves during the night is strictly dominated.
 
\noindent \textbf{Proof.} \\ 
For $n \geq 2m + 1 \geq 7$, we have the following equation from (5):
\begin{equation}
(n - 1 - m) \big[w(n - 2, m) - w(n, m)\big] = m \big[w(n, m) - w(n - 2, m - 1)\big].
\end{equation}
Rearranging the equation, we get:
\begin{equation}
(n - 1) \big[w(n, m) - w(n - 2, m - 1)\big] = (n - 1 - m) \big[w(n - 2, m) - w(n - 2, m - 1)\big].
\end{equation}
Therefore,
\begin{equation}
w(n - 2, m) > w(n, m) \iff w(n, m) > w(n - 2, m - 1) \iff w(n - 2, m) > w(n - 2, m - 1).
\end{equation}

Suppose the werewolf group eliminates one of their own members during the night exactly once, and then proceeds with the ``random strategy+'' afterward. Let $w'(n, m)$ denote their winning probability under this modified strategy. Then:
\begin{equation}
w'(n, m) = \frac{m - 1}{n - 1} w(n - 2, m - 2) + \frac{n - m}{n - 1} w(n - 2, m - 1).
\end{equation}
We now compare \(w(n, m)\) and \(w'(n, m)\):
\begin{equation}
\begin{aligned}
w(n, m) - w'(n, m) &= \left(\frac{n - m - 1}{n - 1}\right) w(n - 2, m) - \left(\frac{m - 1}{n - 1}\right) w(n - 2, m - 2) \\
&\quad + \left(\frac{2m - n}{n - 1}\right) w(n - 2, m - 1).
\end{aligned}
\end{equation}
Since
\begin{equation}
w(n - 2, m) > w(n - 2, m - 1) > w(n - 2, m - 2),
\end{equation}
it follows that
\begin{equation}
w(n, m) - w'(n, m) > 0.
\end{equation}

The remaining cases with smaller values of $n$ or $m$ can be easily verified through direct enumeration.

Therefore, we conclude that for the werewolf group, any strategy involving self-killing during the night is strictly dominated. This completes the justification for the ``random strategy+'' as the optimal strategy.

We now conduct a quantitative analysis of the winning probability of the werewolf group employing the ``random strategy+''.

From the analysis above, we have established that \( w(n,m) > w(n,m-1) \) and \( w(n+2,m) < w(n,m) \). However, the relationship between \( w(n,m) \) and \( w(n-1,m) \) remains undetermined.

To illustrate this, we present a line plot showing the winning probability of the werewolf group in games with 1 to 3 werewolves and up to 20 total players.

\begin{figure}[H]
    \centering
    \includegraphics[width=0.7\textwidth]{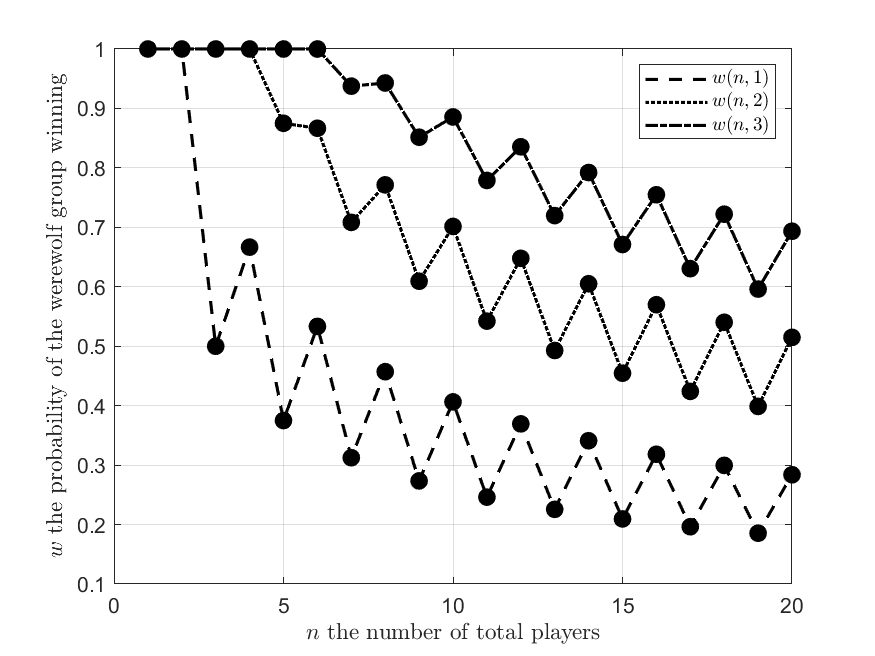}
    \caption{Winning probability of the werwolf group in a game without a prophet}
\end{figure}
 
It is easy to observe from the figure that \( w(2k, m) > w(2k - 1, m) \) when \( k \geq 4 \) and \( k > m \). A similar property also holds under the ``random strategy'', as noted by \textit{Migdał} (2013).

From a game-theoretic perspective, we offer the following intuitive explanation: when the total number of players \( n \) is odd, adding one more citizen (making \( n \) even) does not trigger an additional round of voting. However, this added citizen reduces the per-round probability that a werewolf is voted out. Consequently, for a fixed number of werewolves \( m \), the werewolf group's winning probability increases as the total number of players changes from an odd to the next even number.

Now, we proceed to prove by mathematical induction that for all integers \( k \geq 4 \) and \( m < k \), the inequality \( w(2k, m) > w(2k - 1, m) \) holds.

\subsubsection*{Base Case}

Let \( k = 4 \). We verify the inequality for \( m = 1, 2, 3 \) by explicit computation:
\begin{equation}
w(8, 1) = \frac{48}{105} > w(7, 1) = \frac{5}{16},
\end{equation}
\begin{equation}
w(8, 2) = \frac{27}{35} > w(7, 2) = \frac{39}{55},
\end{equation}
\begin{equation}
w(8, 3) = \frac{33}{35} > w(7, 3) = \frac{15}{16}.
\end{equation}
Thus, the base case holds.

\subsubsection*{Inductive Hypothesis}

Assume that for some \( k \geq 4 \), and for all \( m < k \), we have:
\begin{equation}
w(2k, m) > w(2k - 1, m).
\end{equation}

\subsubsection*{Inductive Step}

We aim to prove:
\begin{equation}
w(2k + 2, m) > w(2k + 1, m), \quad \text{for all } m < k + 1.
\end{equation}

We first show that the inequality holds for all \( m < k \). From the recurrence relations, we have:
\begin{equation}
w(2k + 2, m) = \frac{2k + 1 - m}{2k + 1} w(2k, m) + \frac{m}{2k + 1} w(2k, m - 1),
\end{equation}
\begin{equation}
w(2k + 1, m) = \frac{2k - m}{2k} w(2k - 1, m) + \frac{m}{2k} w(2k - 1, m - 1).
\end{equation}

By the inductive hypothesis:
\begin{equation}
w(2k, m) > w(2k - 1, m) > w(2k - 1, m - 1),
\end{equation}
\begin{equation}
w(2k, m - 1) > w(2k - 1, m - 1).
\end{equation}
Moreover, the coefficients satisfy:
\begin{equation}
\frac{2k + 1 - m}{2k + 1} > \frac{2k - m}{2k}.
\end{equation}
Combining these observations, we conclude:
\begin{equation}
w(2k + 2, m) > w(2k + 1, m), \quad \text{for all } m < k + 1.
\end{equation}

We now consider the boundary case \( m = k \). When \( k \geq 4 \), we have:
\begin{equation}
w(2k + 2, k) = \frac{k + 1}{2k + 1} w(2k, k) + \frac{k}{2k + 1} w(2k, k - 1) = \frac{k + 1}{2k + 1} + \frac{k}{2k + 1} w(2k, k - 1),
\end{equation}
\begin{equation}
w(2k + 1, k) = \frac{1}{2} w(2k - 1, k) + \frac{1}{2} w(2k - 1, k - 1) = \frac{1}{2} + \frac{1}{2} w(2k - 1, k - 1).
\end{equation}
To establish \( w(2k + 2, k) > w(2k + 1, k) \), it suffices to show:
\begin{equation}
w(2k, k - 1) > w(2k - 1, k - 1).
\end{equation}
Since \( w(8, 3) > w(7, 3) \), the argument extends analogously to all larger \( k \).

Therefore, by the principle of mathematical induction, we conclude that for all \( k \geq 4 \) and all \( m < k \), the inequality \( w(2k, m) > w(2k - 1, m) \) holds.

\section{Game with a Prophet}

In this section, we focus on the game with a prophet and aim to derive the optimal strategies for both groups. These strategies collectively induce a \textit{Perfect Bayesian Equilibrium} (PBE).

\subsection{Game under honesty rule}
First, let us consider a simplified but instructive game setting.  
Assume that neither the villagers nor the werewolves can convey false information when announcing messages publicly during the day. We refer to this constraint as the \textit{honesty rule}. Under this rule, it is evident that before the prophet actively reveals the checked information, the ``random strategy~+'' discussed previously is the optimal strategy for both groups. Once the prophet reveals all information obtained from previous nights in a given round, due to the restriction that neither werewolves nor villagers can impersonate the prophet, all players will recognize the identity of the prophet and base their subsequent actions on the revealed information.

For the villagers, this means urging all players to vote out the revealed werewolves.  
Once all revealed werewolves are voted out, villagers will revert to random voting, excluding those who have been verified to be villagers by the prophet. The werewolves, in contrast, would prioritize eliminating the revealed prophet during the first night to prevent further information revelation. Subsequently, they will focus on eliminating the checked villagers during future nights, thereby reducing the likelihood of werewolves being voted out during the day. Naturally, if the number of werewolves equals the number of villagers and one of the werewolves is about to be voted out, the werewolf group will decisively employ the ``all-in strategy'', as they do in the game without a prophet.

\begin{figure}[H]
    \centering
    \includegraphics[width=0.7\textwidth]{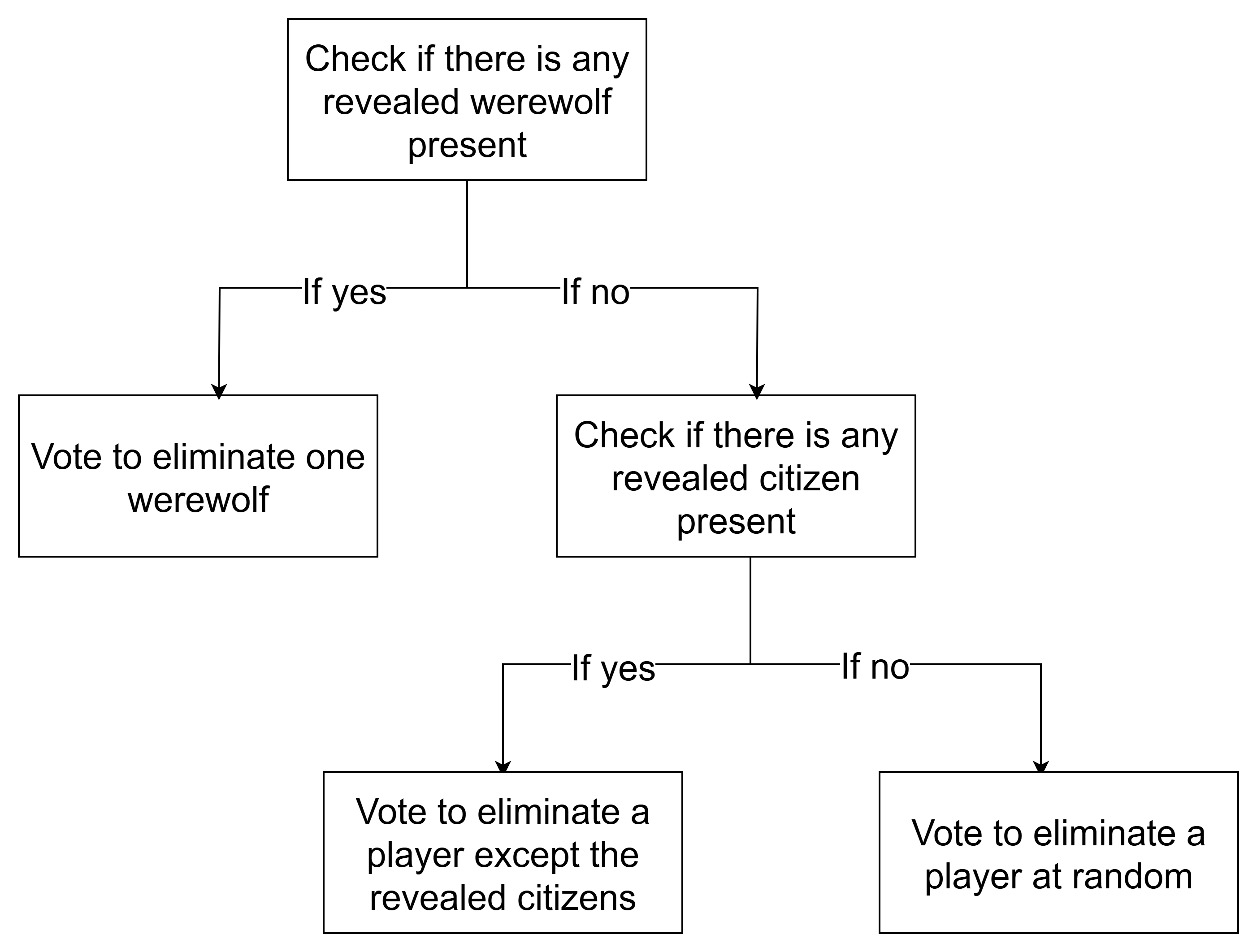}
    \caption{The process of werewolf group choosing the eliminating target}
\end{figure}

\begin{figure}[H]
    \centering
    \includegraphics[width=0.7\textwidth]{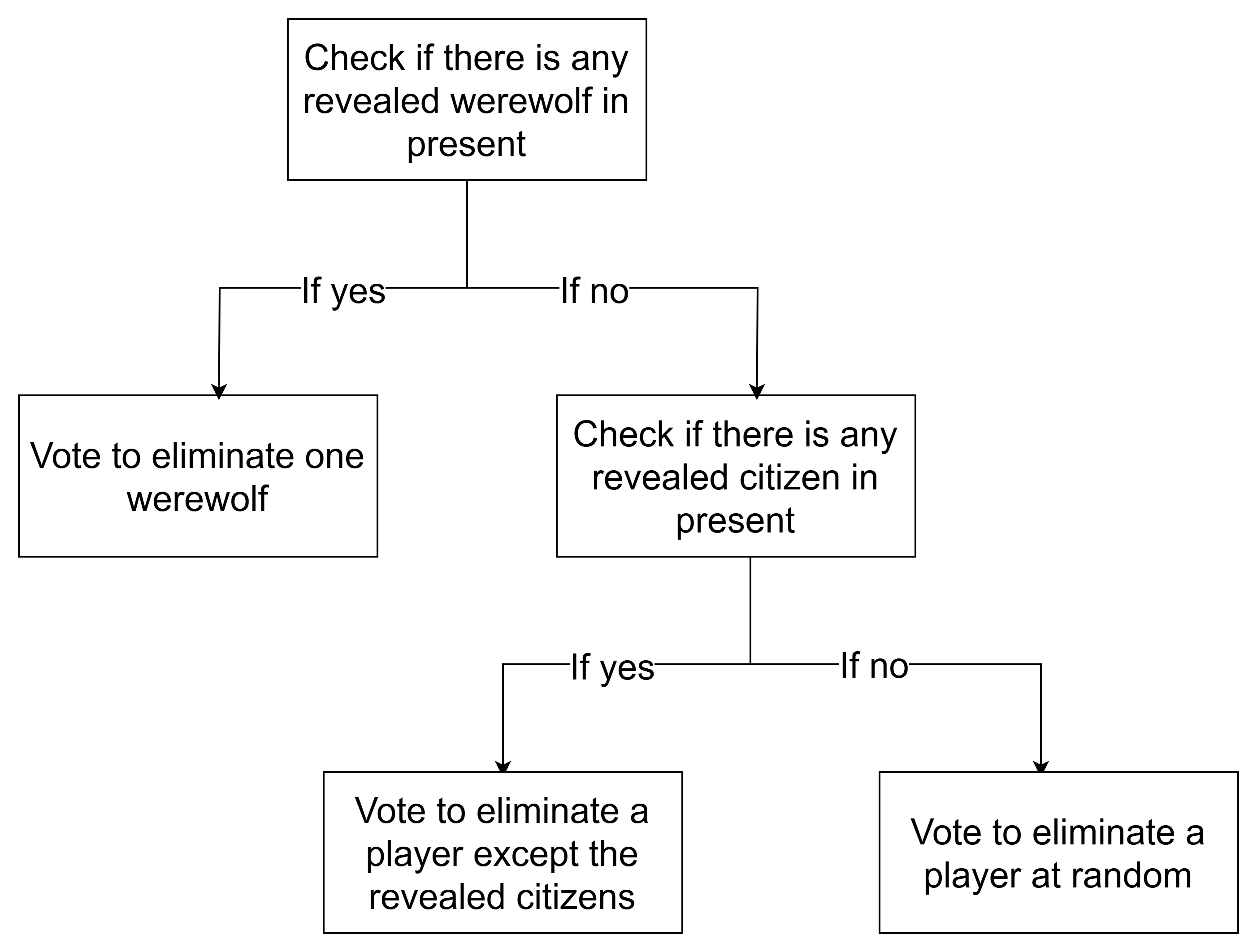}
    \caption{The process of the citizen group choosing the voting target}
\end{figure}

We illustrate the whole game process with a representative example as follows:
\begin{longtable}{|p{16cm}|}
    \caption{One possible process of game with 9 Villagers, 2 Werewolves and 1 Prophet} \\
    \hline
    \endfirsthead

    \hline
    \multicolumn{1}{|c|}{\textbf{(continued)}} \\
    \hline
    \endhead

    \hline
    \endfoot

    \hline
    \endlastfoot

\textbf{Initial role assignment: Player 1: Villager, Player 2: Villager, Player 3: Villager, Player 4: Werewolf, Player 5: Villager, Player 6: Villager, Player 7: Villager, Player 8: Villager, Player 9: Werewolf, Player 10: Villager, Player 11: Villager, Player 12: Prophet} \\
\hline
Night: The Werewolf group eliminated Player 6 (reason: random eliminating citizen) \\
\hline
Remaining Player Profile: 8 Villagers, 2 Werewolves and 1 Prophet \\
\hline
Night: The Prophet checked Player 11, result: Villager \\
\hline
Day: Player 2 was voted out (reason: random voting out) \\
\hline
Remaining Player Profile: 7 Villagers, 2 Werewolves and 1 Prophet \\
\hline
Night: The Werewolf eliminated Player 10 (reason: random eliminating citizen) \\
\hline
Remaining Player Profile: 6 Villagers, 2 Werewolves and 1 Prophet \\
\hline
Night: Player 12, the Prophet checked Player 4, result: Werewolf \\
\hline
Day: The Prophet revealed information: Player 4 is a Werewolf, Player 11 is a Villager \\
\hline
Day: Player 4 was voted out (reason: voting out the revealed werewolf) \\
\hline
Remaining Player Profile: 6 Villagers, 1 Werewolves and 1 Prophet \\
\hline
Night: The Werewolf group killed Player 12 (reason: prioritizing eliminating the revealed prophet) \\
\hline
Remaining Player Profile: 6 Villagers, 1 Werewolves and 0 Prophet \\
\hline
 
Day: Player 1 was voted out (reason: random voting out except the revealed Villager Player 11) \\
\hline
Remaining Player Profile: 5 Villagers, 1 Werewolves and 0 Prophet \\
\hline
Night: The Werewolf eliminated Player 11 (reason: prioritizing eliminating the revealed villager) \\
\hline
Remaining Player Profile: 4 Villagers, 1 Werewolves and 0 Prophet \\
\hline
Day: Player 4 was voted out (reason: random voting out) \\
\hline
Remaining Player Profile: 4 Villagers, 0 Werewolves and 0 Prophet \\
\hline
\textbf{Game Over: The Citizen group won!} \\
\hline
\end{longtable}

In this game, the prophet fortunately checked a werewolf and revealed it in time, which played a crucial role in the final victory of the citizen group.

\subsection{Rule of Thumb in Revealing Information}

The central strategic challenge is to determine the optimal timing for the prophet to reveal their verified information.  
We begin by considering a fixed, \textit{ex-ante} strategy in which the prophet commits in advance to disclosing all known information on the \( x \)-th day of the game.

Given a configuration \((h, m)\), where \( h \) denotes the number of villagers and \( m \) the number of werewolves,  
our objective is to define a mapping that maximizes the expected winning probability of the citizen group when the prophet reveals all checked information on day \( x \).

Let \( H(h, m, x) \) denote the probability that the citizen group wins given \( h \) villagers, \( m \) werewolves, and a prophet who reveals all previously obtained information on day \( x \).  
We define a function \( f: \mathbb{N}^2 \to \mathbb{N} \), where the input \((h, m)\) returns the optimal revelation round \( x \in \mathbb{N} \). Formally:

\[
f(h, m) = \arg \max_{x \in \mathbb{N}} \mathbb{E}[H(h, m, x)]
\]

In essence, the mapping \( f(h, m) \) identifies the round \( x \) that maximizes the citizen group’s ex-ante probability of winning.

To evaluate the ex-ante optimal strategy for the prophet’s revelation, we conduct \textbf{10,000} independent Monte Carlo simulations for each \((h, m)\) configuration.  
For each combination, we estimate the expected citizen winning probability under different revelation rounds and select the round that yields the highest expected probability.  
To account for sampling variability, we compute a 95\% confidence interval (CI) for the citizen winning probability using the standard error of a Bernoulli process. The simulation code is available at: \texttt{https://zenodo.org/records/16366976}

\begin{table}[H]
    \centering
    \caption{Best round $f(h, m)$ of prophet revealing information and the citizen group's winning probability (with 95\% CI)}
    \renewcommand{\arraystretch}{1.3}
    \scriptsize
    \resizebox{\textwidth}{!}{
    \begin{tabular}{|c|c|c|c|c|c|}
        \hline
        \multicolumn{2}{|c|}{} & \multicolumn{4}{c|}{\textbf{Number of werewolves}} \\ \cline{3-6}
        \multicolumn{2}{|c|}{\makecell{\textbf{Citizen Group} \\\ \textbf{Winning Probability}\\\textbf{(95\% CI)}}} & \textbf{1} & \textbf{2} & \textbf{3} & \textbf{4} \\ \hline
        \multirow{9}{*}{\rotatebox{90}{\textbf{Number of villagers}}} 
        & 4  & R1, \textbf{70\%} (68.9, 70.7) & R2, \textbf{37\%} (35.8, 37.6) & R2, \textbf{17\%} (16.3, 17.7) & R1, \textbf{5\%} (4.5, 5.3) \\ \cline{2-6}
        & 5  & R2, \textbf{74\%} (72.5, 74.2) & R2, \textbf{46\%} (45.0, 46.9) & R2, \textbf{22\%} (20.7, 22.4) & R3, \textbf{8\%} (7.9, 8.9) \\ \cline{2-6}
        & 6  & R2, \textbf{77\%} (76.4, 78.1) & R2, \textbf{49\%} (48.3, 50.2) & R3, \textbf{27\%} (26.0, 27.8) & R2, \textbf{12\%} (10.9, 12.1) \\ \cline{2-6}
        & 7  & R2, \textbf{76\%} (75.3, 77.0) & R3, \textbf{52\%} (51.1, 53.1) & R3, \textbf{31\%} (30.3, 32.2) & R3, \textbf{14\%} (13.7, 15.1) \\ \cline{2-6}
        & 8  & R3, \textbf{76\%} (75.4, 77.0) & R3, \textbf{56\%} (55.0, 56.9) & R4, \textbf{33\%} (32.4, 34.3) & R3, \textbf{19\%} (17.9, 19.4) \\ \cline{2-6}
        & 9  & R3, \textbf{78\%} (77.5, 79.1) & R3, \textbf{56\%} (54.7, 56.7) & R3, \textbf{36\%} (35.3, 37.2) & R4, \textbf{23\%} (21.7, 23.4) \\ \cline{2-6}
        & 10 & R3, \textbf{79\%} (78.3, 79.9) & R4, \textbf{58\%} (57.0, 58.9) & R4, \textbf{40\%} (38.5, 40.4) & R4, \textbf{26\%} (24.7, 26.4) \\ \cline{2-6}
        & 11 & R3, \textbf{79\%} (78.2, 79.8) & R4, \textbf{60\%} (58.8, 60.8) & R4, \textbf{42\%} (40.6, 42.5) & R5, \textbf{27\%} (26.4, 28.1) \\ \cline{2-6}
        & 12 & R4, \textbf{80\%} (79.0, 80.6) & R5, \textbf{61\%} (60.2, 62.1) & R5, \textbf{44\%} (42.9, 44.9) & R5, \textbf{30\%} (29.1, 30.9) \\ \hline
    \end{tabular}
    }
\end{table}
  
In standard gameplay, even if the prophet does not strive to follow an interim optimal strategy, simply adhering to the ``rule of thumb'' outlined in the table above significantly increases the citizen group’s likelihood of winning, compared to scenarios where no prophet is present. Here, we provide a comparison between games with and without a prophet, given the same number of citizens.

\begin{figure}[H]
    \centering
    \begin{minipage}[b]{0.45\textwidth}
        \centering
        \includegraphics[width=\textwidth]{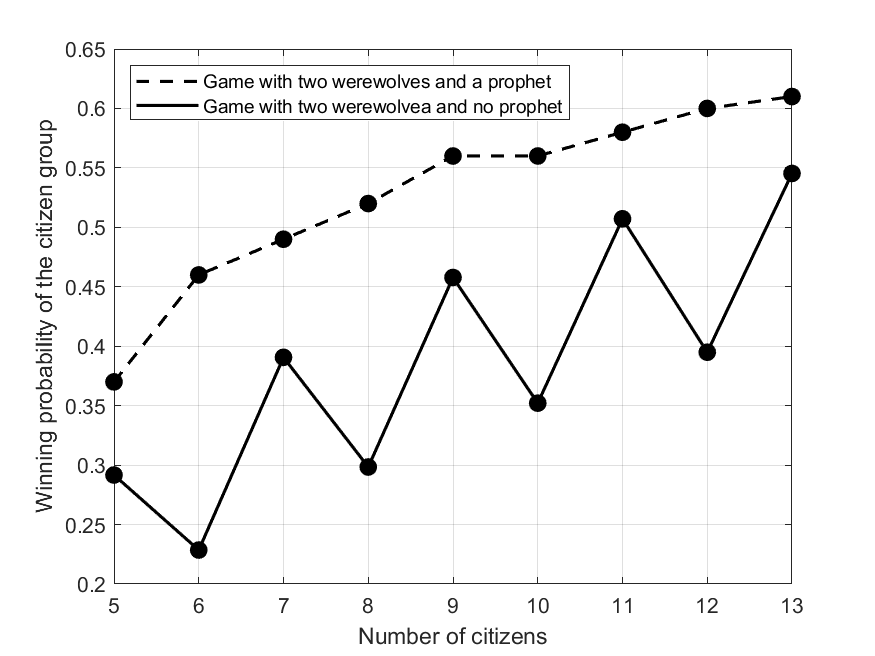}
        \caption{Winning probabilities of the citizen group with and without a prophet in games with two werewolves}
        \label{fig:image1}
    \end{minipage}
    \hspace{0.05\textwidth}  
    \begin{minipage}[b]{0.45\textwidth}
        \centering
        \includegraphics[width=\textwidth]{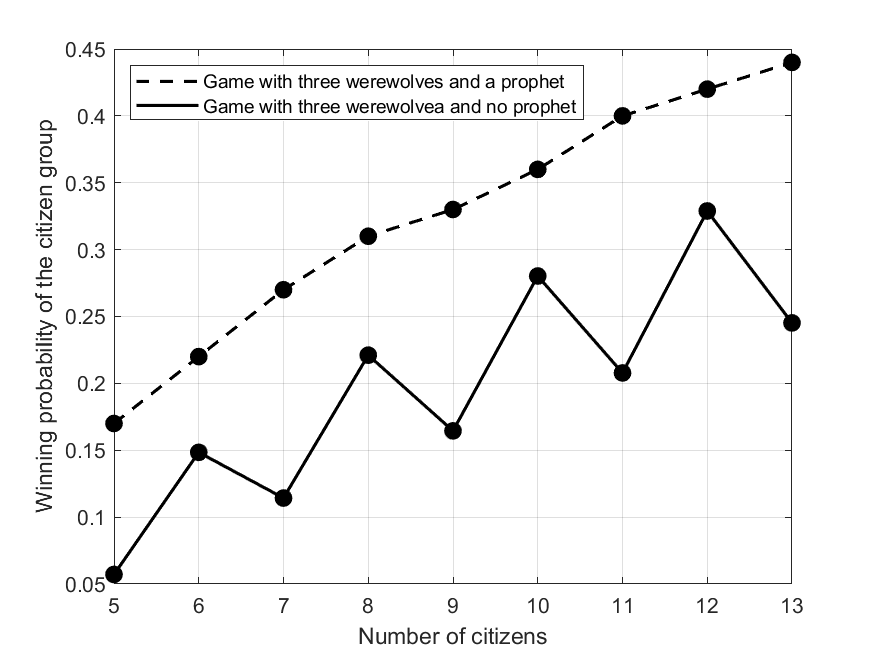}
        \caption{Winning probabilities of the citizen group with and without a prophet in games with three werewolves}
        \label{fig:image2}
    \end{minipage}
\end{figure}

Judging from the comparison results, the presence of a prophet significantly increases the winning probability of the citizen group.

\subsection{PBE under honesty rule}
We aim to construct a strategy that induces a \textit{Perfect Bayesian Equilibrium} (PBE) and maximizes the citizen group's winning probability under the honesty rule. To achieve this, we observe that both before and after the prophet reveals information, all players—including villagers and werewolves—adhere to predetermined strategic patterns as outlined in Section 4.1. Hence, the entire game can be equivalently reformulated as a dynamic game with incomplete information involving only the prophet as an active player. For the prophet, there exist multiple possible role profiles of other players in present (referred to as \textit{nodes}), and the probabilistic distribution over these profiles defines the prophet’s current information set.

Let us denote the information set \( I_t \) possessed by the prophet (if still present) regarding the role profiles of other players at round \( t \) before the message announcement phase as 
\begin{equation}
I_t = \sum_{i} \alpha_i (H_i, M_i, h_i, m_i),
\end{equation}
where
\begin{itemize}
    \item \((H_i, M_i, h_i, m_i)\) represents the different nodes in this information set.
    \item \(H_i\) denotes the number of remaining villagers in node \((H_i, M_i, h_i, m_i)\).
    \item \(M_i\) denotes the number of remaining werewolves in node \((H_i, M_i, h_i, m_i)\).
    \item \(h_i\) denotes the number of checked villagers in node \((H_i, M_i, h_i, m_i)\).
    \item \(m_i\) denotes the number of checked werewolves in node \((H_i, M_i, h_i, m_i)\).
    \item \(\alpha_i\) is the probability of node \((H_i, M_i, h_i, m_i)\) in information set \(I_t\), with \(\sum_i \alpha_i = 1.\)
\end{itemize}

Any two nodes \((H_i, M_i, h_i, m_i)\) and \((H_j, M_j, h_j, m_j)\) within the same information set must satisfy the following conditions:
\begin{align}
H_i + M_i &= H_j + M_j, \\
h_i &= h_j, \\
m_i &= m_j.
\end{align}

For example, suppose a prophet currently holds the information set 
\[
I_t = \frac{1}{3}(5,2,1,1) + \frac{2}{3}(6,1,1,1).
\]
Here, \((5,2,1,1)\) and \((6,1,1,1)\) represent different nodes in information set \(I_t\). This means the prophet knows that among the remaining 7 players, there is a \(\frac{1}{3}\) chance of having 5 villagers and 2 werewolves, and a \(\frac{2}{3}\) chance of having 6 villagers and 1 werewolf. In either case, the prophet has already identified one specific player as a villager and another as a werewolf. 

The requirement that \( h_i = h_j \) and \( m_i = m_j \) holds is straightforward: if the number of confirmed villagers and confirmed werewolves differs across nodes, then the prophet could distinguish between them. As a result, these nodes would not belong to the same information set.

Suppose there exists a mapping \( g : I_t \rightarrow \mathcal{A} \), where \(\mathcal{A} = \{\textit{Hiding}, \textit{Revealing}\}\). Then the prophet’s optimal action is determined by
\begin{equation}
g(I_t) = \arg \max_{x \in \mathcal{A}} R(x, I_t),
\end{equation}
where \(R(x, I_t)\) represents the winning probability of the citizen group if the prophet chooses action \(x\) given the information set \(I_t\). For notational convenience, we denote \(R(g(I_t), I_t) := R(g, I_t)\).

The detailed construction of \( R(\textit{Hiding}, I_t) \) and \( R(\textit{Revealing}, I_t) \) is provided in the Appendix. 

In summary, we have the analytical formula of \(R(\textit{Revealing}, I_t)\):
\begin{equation}
R(\textit{Revealing}, I_t) = s(I_t) = s\left(\sum_i \alpha_i \cdot  (H_i, M_i, h_i, m_i)\right) = \sum_i \alpha_i s(H_i, M_i, h_i, m_i),
\tag{36}
\end{equation}

\begin{equation}
\resizebox{0.9\textwidth}{!}{$
\begin{aligned}
s(H,M,h,m) = 
\begin{cases} 
\frac{M}{H - h + M} \left(\frac{1}{2}\right)^M, & \text{if } M = H + 1 \text{ and } m = 0, \\
\left(\frac{1}{2}\right)^M, & \text{if } M = H + 1 \text{ and } m \geq 1, \\
1 - w(H + M + 2 - 2m, M - m), & \text{if } M < H + 1 \text{ and } m \geq h + 1, \\
u(H + 1 - m, M - m, h + 1 - m), & \text{if } M < H + 1 \text{ and } h + 1 > m \geq 1, \\
\frac{M}{H - h + M} u(H, M - 1, h) + \frac{H - h}{H - h + M} u(H - 1, M, h), & \text{if } M < H + 1 \text{ and } m = 0,
\end{cases}
\end{aligned}
$}
\tag{37}
\end{equation}

where 
\begin{equation}
\resizebox{0.9\textwidth}{!}{$
\begin{aligned}
u(H', M', h') = 
\begin{cases}
1, & \text{if } M' = 0, \\
\frac{H' - h'}{H' + M' - h'} u(H' - 2, M', h' - 1) + \frac{M'}{H' + M' - h'} u(H' - 1, M' - 1, h' - 1), & \text{if } h' \geq 1, M' \geq 1, \text{ and } H' > M', \\
1 - w(H' + M' + 1, M'), & \text{if } h' = 0, M' \geq 1, \text{ and } H' > M', \\
\frac{M'}{H' - h' + M'} \left(\frac{1}{2}\right)^{M'}, & \text{if } H' = M' \text{ and } M' \geq 1, \\
0, & \text{if } H' < M' \text{ and } M' \geq 1.
\end{cases}
\end{aligned}
$}
\tag{38}
\end{equation}

Through recursive computation, all expressions of \( R(\textit{Hiding}, I_t) \) can eventually be written in the following form:
\begin{equation}
R(\textit{Hiding}, I_t) = e(I_t) + \sum_{r} \gamma_{r} R(g, I_{t+1}^{r}),
\tag{39}
\end{equation} 

where \( e(I_t) \) denotes a constant term that differs for each information set \( I_t \), and \( I_{t+1}^{r} \) denotes potential information sets in round \( t+1 \).

In the above discussion, we assumed that the self-killing strategy is not viable for the werewolf group. A rigorous proof is required to demonstrate that our proposed strategy for the prophet, \( g(I_t) \), indeed induces a PBE. \\

\noindent\textbf{Claim 2.} Under the honesty rule, the strategy in which the werewolf group deliberately eliminates one of its own members is strictly dominated.

The proof of Claim 2 is provided in the Appendix.

\subsection{Example}

Now, we give an example in which we solve the strategy for a prophet in a specific information set inducing PBE.

Suppose the prophet is in the information set \(I_t = \frac{2-\sqrt{2}}{2} (5,1,1,1) + \frac{\sqrt{2}}{2} (4,2,1,1).\) In general games, this information set is impossible to exist, but this does not prevent us from calculating the optimal strategy of the prophet under this information set.

First, we calculate \(R(\text{Revealing}, I_t) = s(I_t) = 0.6464\)
through (37) and (38). Then we calculate \(R(\text{Hiding}, I_t) = 0.0918 + \sum_{r=1}^{11} \gamma_{r} R(g, I_{t+1}^{r})\) through (39). The reason why \(r\) goes up to 11 is that one information set constructed by nodes \((3,1,-1,2)\) and \((2,2,-1,2)\) is impossible.\\

\begin{longtable}{|m{0.7cm}|m{1.2cm}|m{6cm}|m{6cm}|}
\caption{Summary of information sets \(I_{t+1}^{r}\)} \\
\hline
\textbf{\(r\)} & \(\boldsymbol{\gamma_{r}}\) & \textbf{Nodes in \(I_{t+1}^{r}\)} & \textbf{Coefficients of nodes in \(I_{t+1}^{r}\)} \\
\hline
\endfirsthead

\hline
\textbf{\(r\)} & \(\boldsymbol{\gamma_{r}}\) & \textbf{Nodes in \(I_{t+1}^{r}\)} & \textbf{Coefficients of nodes in \(I_{t+1}^{r}\)} \\
\hline
\endhead

1 & 0.0273 & (3,1,0,1), (2,2,0,1) & 0.3064, 0.6936 \\
\hline
2 & 0.0630 & (3,1,1,1), (2,2,1,1) & 0.3986, 0.6014 \\
\hline
3 & 0.0189 & (2,2,0,2) & 1 \\
\hline
4 & 0.0443 & (4,0,1,0), (3,1,1,0) & 0.3152, 0.6848 \\
\hline
5 & 0.0734 & (4,0,2,0), (3,1,2,0) & 0.3803, 0.6197 \\
\hline
6 & 0.0050 & (3,1,0,1) & 1 \\
\hline
7 & 0.0152 & (3,1,1,1) & 1 \\
\hline
8 & 0.2084 & (3,1,1,1), (2,2,1,1) & 0.5152, 0.4848 \\
\hline
9 & 0.1579 & (3,1,2,1), (2,2,2,1) & 0.6800, 0.3200 \\
\hline
10 & 0.0505 & (2,2,1,2) & 1 \\
\hline
11 & 0.0253 & (2,2,0,2) & 1 \\
\hline

\end{longtable}

After further calculation, we obtain that 
\[
g(I_{t+1}^{r}) = \textit{Revealing}
\]
for all \( r = 1, 2, \ldots, 11 \). The corresponding values are:
\begin{align*}
R(\textit{Revealing}, I_{t+1}^{1}) &= 0.5916, & R(\textit{Revealing}, I_{t+1}^{2}) &= 0.6993, & R(\textit{Revealing}, I_{t+1}^{3}) &= 1, \\
R(\textit{Revealing}, I_{t+1}^{4}) &= 0.7717, & R(\textit{Revealing}, I_{t+1}^{5}) &= 1, & R(\textit{Revealing}, I_{t+1}^{6}) &= 1, \\
R(\textit{Revealing}, I_{t+1}^{7}) &= 1, & R(\textit{Revealing}, I_{t+1}^{8}) &= 0.7576, & R(\textit{Revealing}, I_{t+1}^{9}) &= 1, \\
R(\textit{Revealing}, I_{t+1}^{10}) &= 1, & R(\textit{Revealing}, I_{t+1}^{11}) &= 1.
\end{align*}

Hence, we compute:
\[
R(\textit{Hiding}, I_t) = 0.0918 + \sum_{ii=1}^{11} \beta_{ii} R(g, I_{t+1}^{ii}) = 0.6903 > R(\textit{Revealing}, I_t) = 0.6464.
\]

Therefore, in the information set \(I_t = \frac{2-\sqrt{2}}{2} (5,1,1,1) + \frac{\sqrt{2}}{2} (4,2,1,1)\), \textit{Hiding} is an optimal strategy.

\section{Conclusion}
This study presents a rigorous analytical framework for modeling the Werewolf Game and opens promising directions for research on decision-making under asymmetric and incomplete information. The mechanisms and equilibria derived—particularly in contexts involving strategic signaling and partial observability—reflect challenges found in real-world domains such as cybersecurity, political negotiation, and competitive markets.

Moreover, the Werewolf Game embodies core features shared by a wide class of social deduction games: asymmetric information, hidden roles, strategic deception, and public communication. As such, the equilibrium analysis and signaling mechanisms developed here are not limited to the Werewolf Game but offer a general modeling framework applicable to other games like \textit{The Resistance}, \textit{Secret Hitler}, and \textit{Among Us}. For instance, roles such as Liberals and Fascists in \textit{Secret Hitler}, or Spies in \textit{The Resistance}, mirror the citizen/werewolf dichotomy in the Werewolf Game. Across these games, strategic communication—often involving limited opportunities for truthful revelation or deliberate misdirection through voting—is a central mechanic. This makes our prophet signaling model broadly applicable in capturing how private information can be credibly conveyed under constraints.

In addition, the analytical tools employed in this study—such as Perfect Bayesian Equilibrium—are suitable for any dynamic game involving evolving information sets and belief-based strategy selection. The algorithm developed to compute optimal strategies for the prophet, grounded in recursive belief updates and decision trees, is also transferable to other roles in similar games. For example, roles like the Engineer or Scientist in \textit{Among Us}, who possess private insights but limited communication opportunities, face decision problems structurally analogous to the prophet's.

Finally, we propose several open questions to guide future research efforts:

\begin{enumerate}
    \item \textbf{Multiple Prophets} \\
    The current analysis focuses primarily on scenarios involving a single prophet. When two or more prophets participate, the game dynamics become substantially more intricate. Key questions arise regarding how the prophets should coordinate their revelation strategies to maximize their collective effectiveness. Furthermore, what would characterize an optimal equilibrium strategy for the citizen group in this multi-prophet setting?
    
    \item \textbf{Absence of Honesty Rule} \\
    Without the honesty rule, players are no longer bound to truthful public communication. This dramatically enlarges the strategic space, as all messages may be deceptive. Even in such more complex or realistic settings, the prophet's decision of when and with what probability to reveal truthful information remains critical. In a sequential speaking framework, the timing and order of communication may significantly influence the outcome. For the citizen group, a natural fallback strategy is to disregard all messages and revert to the ``random strategy+'' employed in prophet-free games. However, this raises several further inquiries:
    \begin{itemize}
        \item Are there alternative strategies that can outperform this fallback approach in terms of citizen winning probability? If such strategies exist, should they be pure or mixed, depending on the player’s role and specific conditions? Intuitively, mixed strategies might provide greater robustness in deceptive environments, but formal analysis is required to substantiate this.
        \item How does the order of speaking affect the information structure and confer strategic advantages to different players?
    \end{itemize}
    
    \item \textbf{Introduction of Additional Roles} \\
    Beyond werewolves, villagers, and prophets, common gameplay often includes other roles such as the guard, who can protect a player from elimination by werewolves during the night. How do these additional roles alter the strategic landscape and dynamics of the game?
\end{enumerate}

\section*{Declarations}
\subsection*{Funding and/or Conflicts of interests}
The author did not receive support from any organization for the submitted work. The author has no relevant financial or non-financial interests to disclose.
\subsection*{Data availability}
All data we demonstrate and analyze is generated through theoretical and mathematical approaches in this papaer.

\section*{Appendix}
\subsection*{Construction of $R(x, I_t) $}
Denote \(R(\textit{Revealing}, I_t) = s(I_t)\). After the prophet reveals the information, the winning probability of the citizen group is the weighted sum of winning probabilities in each node, because the same action patterns are adopted by both the citizen group and the werewolf group:
\begin{equation}
s(I_t) = s\left(\sum_i \alpha_i \cdot (H_i, M_i, h_i, m_i)\right) = \sum_i \alpha_i s(H_i, M_i, h_i, m_i).
\end{equation}

We consider two different cases of \( s(H,M,h,m) \), namely, when the number of werewolves that have been checked is greater than or equal to the number of villagers that have been checked plus the number of prophets, and when the number of werewolves that have been checked is smaller than the number of villagers that have been checked plus the number of prophets. The former case is relatively simple and can be directly transformed into the form of the game without a prophet. The latter case is a little more complex, in which there exist some checked villagers and no checked werewolves after some rounds processed by the action patterns adopted by both groups. Then we obtain the recursive formula:
\begin{equation}
\resizebox{0.9\textwidth}{!}{$
\begin{aligned}
s(H,M,h,m) = 
\begin{cases} 
\frac{M}{H - h + M} \left(\frac{1}{2}\right)^{M}, & \text{if } M = H + 1 \text{ and } m = 0, \\
\left(\frac{1}{2}\right)^{M}, & \text{if } M = H + 1 \text{ and } m \geq 1, \\
1 - w(H + M + 2 - 2m, M - m), & \text{if } M < H + 1 \text{ and } m \geq h + 1, \\
u(H + 1 - m, M - m, h + 1 - m), & \text{if } M < H + 1 \text{ and } h + 1 > m \geq 1, \\
\frac{M}{H - h + M} u(H, M - 1, h) + \frac{H - h}{H - h + M} u(H - 1, M, h), & \text{if } M < H + 1 \text{ and } m = 0.
\end{cases} 
\end{aligned}
$}
\end{equation}

where
\begin{equation}
\resizebox{0.9\textwidth}{!}{$
\begin{aligned}
u(H', M', h') = 
\begin{cases}
1, & \text{if } M' = 0, \\
\frac{H' - h'}{H' + M' - h'} u(H' - 2, M', h' - 1) + \frac{M'}{H' + M' - h'} u(H' - 1, M' - 1, h' - 1), & \text{if } h' \geq 1, M' \geq 1, \text{ and } H' > M', \\
1 - w(H' + M' + 1, M'), & \text{if } h' = 0, M' \geq 1, \text{ and } H' > M', \\
\frac{M'}{H' - h' + M'} \left(\frac{1}{2}\right)^{M'}, & \text{if } H' = M' \text{ and } M' \geq 1, \\
0, & \text{if } H' < M' \text{ and } M' \geq 1.
\end{cases}
\end{aligned}
$}
\end{equation}

The function \( u(H', M', h') \) characterizes the behavior of both groups after the prophet reveals information identifying \( h' \) villagers, with no werewolves exposed. Here, \( H' \) denotes the total number of villagers, \( M' \) the total number of werewolves, and \( h' \) the number of villagers whose group affiliations have been verified by the prophet.

This function reflects the citizen group’s strategy to avoid voting out the confirmed villagers during the day, while the werewolf group, in turn, prioritizes eliminating these revealed villagers during the night.

Now, we give the recursive formula of $R(Hiding, I_t)$.  
\begin{equation}
\begin{aligned}
R(Hiding, I_t) = & \sum_{i} \alpha_i P_{17}(H_i, M_i, h_i, m_i) (1 - w(H_i + M_i, M_i)) \\
& + \sum_{i} \alpha_i P_{18}(H_i, M_i, h_i, m_i) (1 - w(H_i + M_i, M_i)) \\
& + \sum_{i} \alpha_i P_{19}(H_i, M_i, h_i, m_i) (1 - w(H_i + M_i, M_i - 1)) \\
& + \sum_{i} \alpha_i P_{20}(H_i, M_i, h_i, m_i) (1 - w(H_i + M_i, M_i - 1)) \\
& + \sum_{i} \alpha_i P_{21}(H_i, M_i, h_i, m_i) (1 - w(H_i + M_i, M_i)) \\
& + \sum_{ii=1}^{8}\left( \sum_{i} \alpha_i P_{ii}(H_i, M_i, h_i, m_i) 
R\left(g, \sum_{i} \frac{\alpha_i P_{ii}(H_i, M_i, h_i, m_i) node_{ii}(H_i, M_i, h_i, m_i)}{\sum_{i} \alpha_i P_{ii}(H_i, M_i, h_i, m_i)} \right) \right)  \\
& + \sum_{ii=9}^{12} \Bigg( \sum_{i} \alpha_i \Big( P_{ii}(H_i, M_i, h_i, m_i) + P_{ii+4}(H_i, M_i, h_i, m_i) \Big) \cdot \\
& \quad R\Bigg(g, \sum_{i} \frac{\alpha_i P_{ii}(H_i, M_i, h_i, m_i) \cdot node_{ii}(H_i, M_i, h_i, m_i)}{\sum_{i} \alpha_i \Big( P_{ii}(H_i, M_i, h_i, m_i) + P_{ii+4}(H_i, M_i, h_i, m_i) \Big)} \\
& \quad + \sum_{i} \frac{\alpha_i P_{ii+4}(H_i, M_i, h_i, m_i) \cdot node_{ii+4}(H_i, M_i, h_i, m_i)}{\sum_{i} \alpha_i \Big( P_{ii}(H_i, M_i, h_i, m_i) + P_{ii+4}(H_i, M_i, h_i, m_i) \Big)} \Bigg) \Bigg).
\end{aligned}
\end{equation}
where\\
\(P_{ii}\) denotes the probability of \((H,M,h,m)\) turning into \(node_{ii}(H,M,h,m)\).\\
\(P_{17}\) denotes situation where prophet is voted out during the day.\\
\(P_{18}\) denotes situation where checked villager is voted out during the day and prophet is eliminated during the night.\\
\(P_{19}\) denotes situation where checked werewolf is voted out during the day and prophet is eliminated during the night.\\
\(P_{20}\) denotes situation where unchecked werewolf is voted out during the day and prophet is eliminated during the night.\\
\(P_{21}\) denotes situation where unchecked villager is voted out during the day and prophet is eliminated during the night.\\
\(node_{1}(H,M,h,m)\) is induced by Checked villager voted out-Villager checked-Checked villager eliminated;\\
\(node_{2}(H,M,h,m)\) is induced by Checked villager voted out-Villager checked-Unchecked villager eliminated;\\
\(node_{3}(H,M,h,m)\) is induced by Checked villager voted out-Werewolf checked-Checked villager eliminated;\\
\(node_{4}(H,M,h,m)\) is induced by Checked villager voted out-Werewolf checked-Unchecked villager eliminated;\\
\(node_{5}(H,M,h,m)\) is induced by Checked werewolf voted out-Villager checked-Checked villager eliminated;\\
\(node_{6}(H,M,h,m)\) is induced by Checked werewolf voted out-Villager checked-Unchecked villager eliminated;\\
\(node_{7}(H,M,h,m)\) is induced by Checked werewolf voted out-Werewolf checked-Checked villager eliminated;\\
\(node_{8}(H,M,h,m)\) is induced by Checked werewolf voted out-Werewolf checked-Unchecked villager eliminated;\\
\(node_{9}(H,M,h,m)\) is induced by Unchecked werewolf voted out-Villager checked-Checked villager eliminated;\\
\(node_{10}(H,M,h,m)\) is induced by Unchecked werewolf voted out-Villager checked-Unchecked villager eliminated;\\
\(node_{11}(H,M,h,m)\) is induced by Unchecked werewolf voted out-Werewolf checked-Unchecked villager eliminated;\\
\(node_{12}(H,M,h,m)\) is induced by Unchecked werewolf voted out-Werewolf checked-Checked villager eliminated;\\
\(node_{13}(H,M,h,m)\) is induced by Unchecked villager voted out-Villager checked-Checked villager eliminated;\\
\(node_{14}(H,M,h,m)\) is induced by Unchecked villager voted out-Villager checked-Unchecked villager eliminated;\\
\(node_{15}(H,M,h,m)\) is induced by Unchecked villager voted out-Werewolf checked-Unchecked villager eliminated;\\
\(node_{16}(H,M,h,m)\) is induced by Unchecked villager voted out-Werewolf checked-Checked villager eliminated.\\

The specific expressions of the variable are as follows:
\\
\begin{equation}
P_1(H_i, M_i, h_i, m_i) = \frac{h_i}{H_i + M_i + 1} \cdot \frac{(H_i - 1) - (h_i - 1)}{(H_i - 1) + M_i - (h_i - 1) - m_i} \cdot \frac{h_i}{H_i}
\end{equation}

\begin{equation}
P_2(H_i, M_i, h_i, m_i) = \frac{h_i}{H_i + M_i + 1} \cdot \frac{(H_i - 1) - (h_i - 1)}{(H_i - 1) + M_i - (h_i - 1) - m_i} \cdot \frac{H_i - 1 - h_i}{H_i}
\end{equation}

\begin{equation}
P_3(H_i, M_i, h_i, m_i) =  \frac{h_i}{H_i + M_i + 1} \cdot \frac{M_i - m_i}{(H_i - 1) + M_i - (h_i - 1) - m_i} \cdot \frac{h_i - 1}{H_i}  
\end{equation}

\begin{equation}
\small
P_4(H_i, M_i, h_i, m_i) =  \frac{h_i}{H_i + M_i + 1} \cdot \frac{M_i - m_i}{(H_i - 1) + M_i - (h_i - 1) - m_i} \cdot \frac{(H_i - 1) - (h_i - 1)}{H_i}
\end{equation}

\begin{equation}
P_5(H_i, M_i, h_i, m_i) = \frac{m_i}{H_i + M_i + 1} \cdot \frac{H_i-h_i}{H_i + (M_i - 1) - h_i - (m_i - 1)} \cdot \frac{h_i + 1}{H_i + 1}
\end{equation}

\begin{equation}
P_6(H_i, M_i, h_i, m_i) = \frac{m_i}{H_i + M_i + 1} \cdot \frac{H_i-h_i}{H_i + (M_i - 1) - h_i - (m_i - 1)} \cdot \frac{H_i - h_i}{H_i + 1}
\end{equation}

\begin{equation}
P_7(H_i, M_i, h_i, m_i) = \frac{m_i}{H_i + M_i + 1} \cdot \frac{(M_i - 1) - (m_i - 1)}{H_i + (M_i - 1) - h_i - (m_i - 1)} \cdot \frac{h_i}{H_i + 1}
\end{equation}

\begin{equation}
P_8(H_i, M_i, h_i, m_i) = \frac{m_i}{H_i + M_i + 1} \cdot \frac{(M_i - 1) - (m_i - 1)}{H_i + (M_i - 1) - h_i - (m_i - 1)} \cdot \frac{H_i - h_i}{H_i + 1}
\end{equation} 

\begin{equation}
P_9(H_i, M_i, h_i, m_i) = \frac{M_i - m_i}{H_i + M_i + 1} \cdot \frac{H_i - h_i}{H_i + (M_i - 1) - h_i - m_i} \cdot \frac{h_i + 1}{H_i + 1}
\end{equation}

\begin{equation}
P_{10}(H_i, M_i, h_i, m_i) = \frac{M_i - m_i}{H_i + M_i + 1} \cdot \frac{H_i - h_i}{H_i + (M_i - 1) - h_i - m_i} \cdot \frac{H_i - (h_i+1)}{H_i + 1}
\end{equation}

\begin{equation}
P_{11}(H_i, M_i, h_i, m_i) = \frac{M_i - m_i}{H_i + M_i + 1} \cdot \frac{(M_i-1)- m_i}{H_i + (M_i - 1) - h_i - m_i} \cdot \frac{H_i - h_i}{H_i + 1}
\end{equation}

\begin{equation}
P_{12}(H_i, M_i, h_i, m_i) = \frac{M_i - m_i}{H_i + M_i + 1} \cdot \frac{(M_i - 1) - m_i}{H_i + (M_i - 1) - h_i - m_i} \cdot \frac{h_i}{H_i + 1}
\end{equation}

\begin{equation}
P_{13}(H_i, M_i, h_i, m_i) = \frac{H_i - h_i}{H_i + M_i + 1} \cdot \frac{(H_i - 1) - h_i}{(H_i - 1) + M_i - h_i - m_i} \cdot \frac{h_i + 1}{H_i}
\end{equation}

\begin{equation}
P_{14}(H_i,M_i,h_i,m_i) = \frac{H_i-h_i}{H_i + M_i + 1} \cdot \frac{(H_i-1)-h_i}{(H_i-1)+M_i  -h_i-m_i} \cdot \frac{(H_i-1)-(h_i+1)}{H_i}
\end{equation}

\begin{equation}
P_{15}(H_i,M_i,h_i,m_i) = \frac{H_i-h_i}{H_i + M_i + 1} \cdot \frac{M_i-m_i}{(H_i-1)+M_i  -h_i-m_i} \cdot \frac{(H_i-1)-h_i}{H_i}
\end{equation}

\begin{equation}
P_{16}(H_i,M_i,h_i,m_i) = \frac{H_i-h_i}{H_i + M_i + 1} \cdot \frac{M_i-m_i}{(H_i-1)+M_i  -h_i-m_i} \cdot \frac{h_i}{H_i}
\end{equation}

\begin{equation}
P_{17}(H_i, M_i, h_i, m_i) = \frac{1}{H_i + M_i + 1}
\end{equation}

\begin{equation}
P_{18}(H_i, M_i, h_i, m_i) = \frac{h_i}{H_i + M_i + 1} \cdot \frac{1}{H_i-1 + 1}
\end{equation}

\begin{equation}
P_{19}(H_i, M_i, h_i, m_i) = \frac{m_i}{H_i + M_i + 1} \cdot \frac{1}{H_i + 1}
\end{equation}

\begin{equation}
P_{20}(H_i, M_i, h_i, m_i) = \frac{M_i-m_i}{H_i + M_i + 1} \cdot \frac{1}{H_i + 1}
\end{equation}

\begin{equation}
P_{21}(H_i, M_i, h_i, m_i) = \frac{H_i-h_i}{H_i + M_i + 1} \cdot \frac{1}{H_i-1 + 1}
\end{equation}

\begin{equation}
node_1(H_i, M_i, h_i, m_i) = (H_i-2,M_i,h_i-1,m_i)
\end{equation}

\begin{equation}
node_2(H_i, M_i, h_i, m_i) = (H_i-2,M_i,h_i,m_i)
\end{equation}

\begin{equation}
node_3(H_i, M_i, h_i, m_i) = (H_i-2,M_i,h_i-2,m_i+1)
\end{equation}

\begin{equation}
node_4(H_i, M_i, h_i, m_i) = (H_i-2,M_i,h_i-1,m_i+1)
\end{equation}

\begin{equation}
node_5(H_i, M_i, h_i, m_i) = (H_i-1,M_i-1,h_i,m_i-1)
\end{equation}

\begin{equation}
node_6(H_i, M_i, h_i, m_i) = (H_i-1,M_i-1,h_i+1,m_i-1)
\end{equation}

\begin{equation}
node_7(H_i, M_i, h_i, m_i) = (H_i-1,M_i-1,h_i-1,m_i)
\end{equation}

\begin{equation}
node_8(H_i, M_i, h_i, m_i)= (H_i-1,M_i-1,h_i,m_i)
\end{equation}

\begin{equation}
node_9(H_i, M_i, h_i, m_i) = (H_i-1,M_i-1,h_i,m_i)
\end{equation}

\begin{equation}
node_{10}(H_i, M_i, h_i, m_i) = (H_i-1,M_i-1,h_i+1,m_i)
\end{equation}

\begin{equation}
node_{11}(H_i, M_i, h_i, m_i) = (H_i-1,M_i-1,h_i,m_i+1)
\end{equation}

\begin{equation}
node_{12}(H_i, M_i, h_i, m_i) = (H_i-1,M_i-1,h_i-1,m_i+1)
\end{equation}

\begin{equation}
node_{13}(H_i, M_i, h_i, m_i) = (H_i-2,M_i,h_i,m_i)
\end{equation}

\begin{equation}
node_{14}(H_i, M_i, h_i, m_i) = (H_i-2,M_i,h_i+1,m_i)
\end{equation}

\begin{equation}
node_{15}(H_i, M_i, h_i, m_i) = (H_i-2,M_i,h_i,m_i+1)
\end{equation}

\begin{equation}
node_{16}(H_i, M_i, h_i, m_i) = (H_i-2,M_i,h_i-1,m_i+1)
\end{equation}

\begin{equation}
node_{17}(H_i, M_i, h_i, m_i) = (H_i-2,M_i,h_i,m_i)
\end{equation}

\begin{equation}
node_{18}(H_i, M_i, h_i, m_i) = (H_i,M_i,h_i,m_i+1)
\end{equation}

Now we introduce the transformation criteria for information sets. If the current information set \( I_t \) satisfies any of the following conditions, it should first be transformed into an equivalent form, as defined below, before being substituted into (39).

If no valid nodes remain within the information set after applying this transformation, the prophet’s decision-making process is considered complete. The game effectively terminates for the prophet.

\begin{enumerate}
    \item There exists some node in the information set \( I_t \) such that the werewolf group wins directly.
    
    For the information set \( I_t = \sum_i \alpha_i \cdot (H_i, M_i, h_i, m_i) \), if there exists any \( i \) such that \( H_i + 1 < M_i \), let \( J \) be the set of such indices. Then we rewrite:
    \begin{equation}
        R(\textit{Hiding}, I_t) = \left( \sum_{i \notin J} \alpha_i \right) R\left( \textit{Hiding}, \sum_{i \notin J} \frac{\alpha_i \cdot (H_i, M_i, h_i, m_i)}{\sum_{i \notin J} \alpha_i} \right).
    \end{equation}
    Naturally, if the prophet instead chooses the \textit{Revealing} strategy, the werewolf group will still win in these nodes, which is already captured in \( R(\textit{Revealing}, I_t) \).

    \item There exists some node in the information set \( I_t \) such that the citizen group wins directly.
    
    For the information set \( I_t = \sum_i \alpha_i \cdot (H_i, M_i, h_i, m_i) \), if there exists any \( i \) such that \( M_i = 0 \), let \( K \) be the set of such indices. Then we rewrite:
    \begin{equation}
        R(\textit{Hiding}, I_t) = \sum_{i \in K} \alpha_i + \left( \sum_{i \notin K} \alpha_i \right) R\left( \textit{Hiding}, \sum_{i \notin K} \frac{\alpha_i \cdot (H_i, M_i, h_i, m_i)}{\sum_{i \notin K} \alpha_i} \right).
    \end{equation}

    \item There exists some node in the information set \( I_t \) such that the werewolf group would directly employ the ``all-in strategy''.
    
    For the information set \( I_t = \sum_i \alpha_i \cdot (H_i, M_i, h_i, m_i) \), if there exists any \( i \) such that \( H_i + 1 = M_i \), let \( L \) be the set of such indices. Then we rewrite:
    \begin{equation}
        R(\textit{Hiding}, I_t) = \sum_{i \in L} \alpha_i \left(\frac{1}{2}\right)^{M_i + 1} + \sum_{i \notin L} \alpha_i R\left( \textit{Hiding}, \sum_{i \notin L} \frac{\alpha_i \cdot (H_i, M_i, h_i, m_i)}{\sum_{i \notin L} \alpha_i} \right).
    \end{equation}
\end{enumerate}

Combining the recursive formulation of \( R(\textit{Hiding}, I_t) \) with the transformation criteria discussed above, we observe that each information set \( I_t \) can generate at most twelve successor information sets at time \( t+1 \). Denote these potential information sets by \( \{ I_{t+1}^r \}_r \). Then, (39) can be written as
\begin{equation}
R(\textit{Hiding}, I_t) = e(I_t) + \sum_{r} \gamma_r R(g, I_{t+1}^r),
\tag{39}
\end{equation}
where \( e(I_t) \) denotes a constant term that differs for each information set \( I_t \).

The value \( R(g, I_t) \) is defined as the maximum between \( R(\textit{Revealing}, I_t) \) and \( R(\textit{Hiding}, I_t) \). As shown in (39), the value of \( R(\textit{Hiding}, I_t) \) is a linear combination of all \( R(g, I_{t+1}^r) \) for \( I_{t+1}^r \in \{ I_{t+1}^r \}_r \). For each \( R(g, I_{t+1}^r) \), we apply the same process recursively, just as we did for \( R(g, I_t) \).

\begin{figure}[H]
    \centering
    \includegraphics[width=\textwidth]{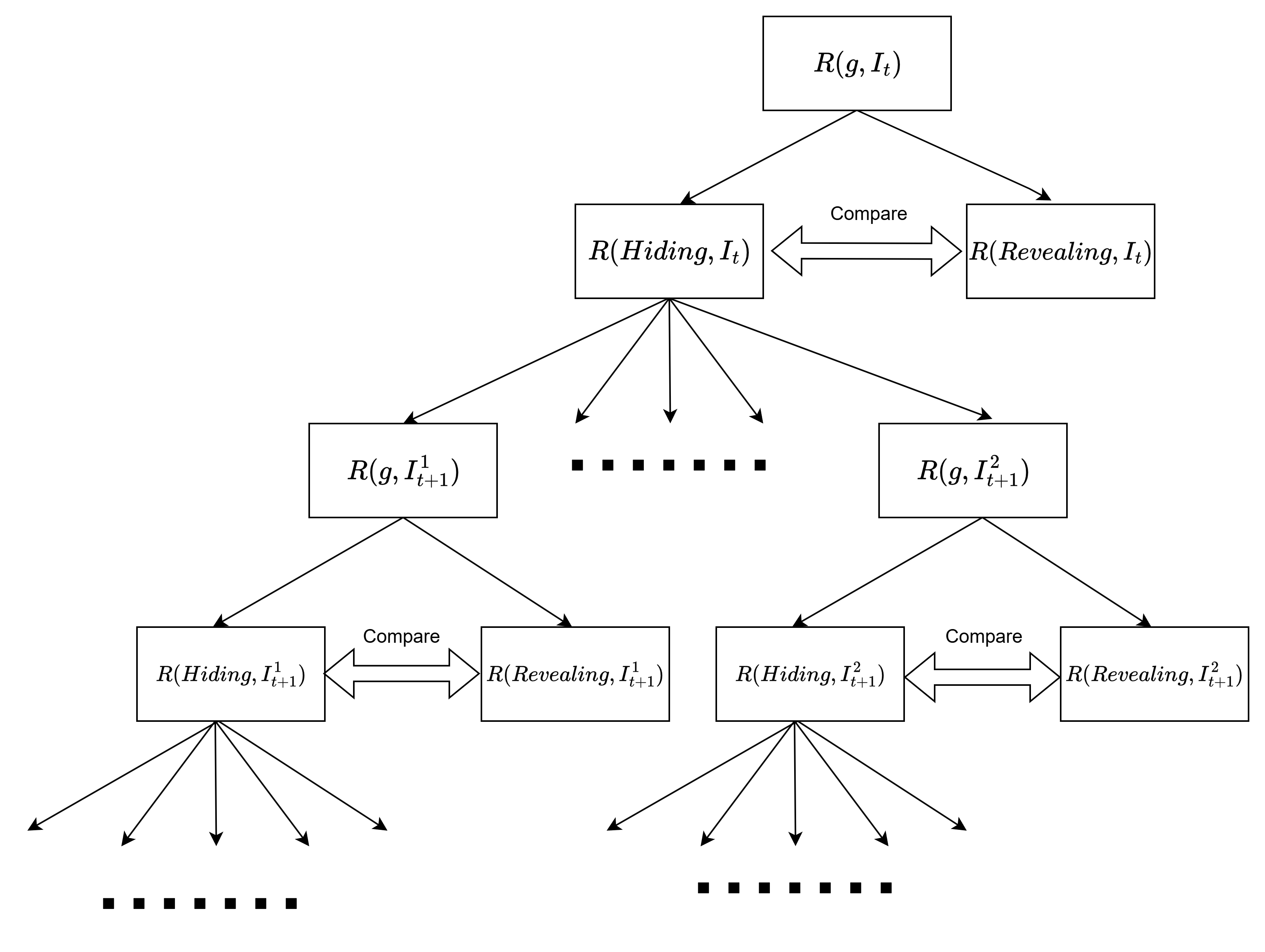}
    \caption{Flowchart of calculating \( R(\textit{Hiding}, I_t) \).}
\end{figure}

Through dynamic programming, we can calculate the prophet's optimal strategy given any information set in the game. Moreover, the entire process can be regarded as a \textit{Markov Decision Process} (MDP), meaning that no matter how the prophet arrives at a certain information set, as long as the information set is the same, the optimal action should also be the same. We can even derive optimal actions in certain information sets that cannot be reached through regular gameplay, such as those involving irrational node probabilities. The entire proc

\subsection*{Proof of Claim 2}
\textbf{Proof.}  

First, consider the scenario where the werewolf group might self-kill during the night, which can be divided into two cases:
\begin{itemize} 
\item The prophet has revealed the checking results; 
\item The prophet has not yet revealed the checking results.
\end{itemize}

The case where the prophet has revealed their information is straightforward to analyze. Once the prophet has revealed the information set \( I_t = \sum \alpha_i (H_i,M_i,h_i,m_i) \), the winning probability of the citizen group is \(\sum_i \alpha_i s(H_i, M_i, h_i, m_i)\). For any \( s(H_i, M_i, h_i, m_i) \), it can be regarded as a variation of the game without prophets. Therefore, any strategy involving self-killing for the werewolf group is dominated in this case.

The case where the prophet has not yet revealed their information can be further divided into two subcases:
\begin{itemize} 
\item The prophet has been eliminated either by werewolf elimination or by being voted out; 
\item The prophet is still present.
\end{itemize}

If the prophet has been eliminated, the game reduces to one without a prophet. In this situation, the werewolf group self-killing during the night is a strictly dominated strategy, which has been proven in Section 3.

If the prophet is still present, the situation becomes more complex, and we provide a detailed proof below.

Given the complexity of the recursive model introduced earlier, along with its multiple termination conditions, directly analyzing the potential payoffs for the prophet across various information sets is challenging. To address this, we adopt a \textit{Dynamic Adjustment Process} approach, similar to that used in the Cournot duopoly model, where both parties continuously adapt to each other's strategies, iteratively adjusting their actions until convergence to an equilibrium.

Suppose the information set of the prophet after the checking and before the werewolf elimination is \(\sum_{i=1}^n \beta_i \cdot (H_i,M_i,h,m)\). Assume that in the prophet's belief, the probability of the werewolf group adopting a self-killing strategy at node \((H_i, M_i, h, m)\) is \(\hat{q}_i\). Let the vector \(\hat{Q} = [\hat{q}_1, \dots, \hat{q}_n]\) represent these independent probabilities. If the actual probability vector of the werewolf group's self-killing actions equals \(\hat{Q}\), we denote the expected winning probability of the citizen group under strategy profile \(\hat{Q}\) as \(AR(\hat{Q})\). Formally, this can be expressed as:

\begin{align}
AR(\hat{Q}) = & \sum_i \beta_i \hat{q}_i \frac{m}{M_i} R\left(g, \sum_i \frac{\beta_i \hat{q}_i \frac{m}{M_i} (H_i, M_i-1, h, m-1)}{\sum_i \beta_i \hat{q}_i \frac{m}{M_i}}\right) \notag \\
& + \sum_i \beta_i \left(\hat{q}_i \frac{M_i - m}{M_i} + (1-\hat{q}_i) \frac{H_i - h}{H_i + 1}\right) \notag \\
& \quad \cdot R\left(g, \sum_i \frac{\beta_i \left(\hat{q}_i \frac{M_i - m}{M_i} (H_i, M_i - 1, h, m) + (1 - \hat{q}_i) \frac{H_i - h}{H_i + 1} (H_i - 1, M_i, h, m)\right)}{\sum_i \beta_i \left(\hat{q}_i \frac{M_i - m}{M_i} + (1 - \hat{q}_i) \frac{H_i - h}{H_i + 1}\right)} \right) \notag \\
& + \sum_i \beta_i (1 - \hat{q}_i) \frac{1}{H_i + 1} \left(1 - w(H_i + M_i + 1, M_i)\right) \notag \\
& + \sum_i \beta_i (1 - \hat{q}_i) \frac{h}{H_i + 1} R\left(g, \sum_i \frac{\beta_i (1 - \hat{q}_i) \frac{h}{H_i + 1} (H_i - 1, M_i, h - 1, m)}{\sum_i \beta_i (1 - \hat{q}_i) \frac{h}{H_i + 1}} \right)
\end{align}

It is straightforward to see that \( AR(\hat{Q}) \) is monotonically increasing with respect to each component of \(\hat{Q}\). This implies that when the probability distribution of the werewolf group's actions is fully known to the prophet, self-killing strategies cannot mislead the prophet’s judgment. Instead, such strategies only decrease the werewolf group’s winning probability.

Suppose the actual probability of the werewolf group self-killing is \( Q = [q_1, \dots, q_n] \) and the actual expectation of the citizen group’s winning probability, when the prophet takes the optimal strategy given belief \(\hat{Q}\), is \( DR(Q, \hat{Q}) \). When \( Q \leq \hat{Q} \), we have

\begin{equation}
DR(Q, \hat{Q}) \leq AR(Q) \leq AR(\hat{Q}),
\end{equation}
with equality holding if and only if \( Q = \hat{Q} \).

From the perspective of the werewolf group, given any \(\hat{Q}\) from the prophet, they could choose \( Q < \hat{Q} \) to reduce the citizen group’s winning probability. Similarly, from the prophet's perspective, given any \( Q \) from the werewolf group, the prophet can ensure \(\hat{Q} = Q\). Thus, we conclude \( Q = \hat{Q} = \vec{0} \), meaning that in any case, the probability of the werewolf group adopting a self-killing strategy is zero.

\textbf{Proof complete.}


\begin{thebibliography}{99}  
\bibitem{WikipediaMafia}
    Wikipedia. ``Mafia (Party Game)." \textit{Wikipedia, The Free Encyclopedia}, 1 June 2024. [Online]. Available: \url{https://en.wikipedia.org/wiki/Mafia_(party_game)}. [Accessed: 1 June 2024].

\bibitem{Braverman2008}
Braverman, Mark, Omid Etesami, and Elchanan Mossel. ``Mafia: A Theoretical Study of Players and Coalitions in a Partial Information Environment.'' \textit{Proceedings of the 3rd International Conference on Algorithmic Game Theory}, 2008, pp. 825-846.

\bibitem{Yao2008}
Yao, Erlin. ``A Theoretical Study of Mafia Games.'' \textit{arXiv preprint arXiv:0804.0071}, 2008.

\bibitem{Migdal2010}
Migdał, Piotr. ``A Mathematical Model of the Mafia Game.'' \textit{arXiv preprint arXiv:1009.1031}, 2010.

\bibitem{Bi2016}
Bi, Xiaoheng, and Tetsuro Tanaka. ``Human-Side Strategies in the Werewolf Game Against the Stealth Werewolf Strategy.'' \textit{International Conference on Computers and Games}. Cham: \textit{Springer International Publishing}, 2016.

\bibitem{Xiong2017}
Xiong, Shuo, et al. ``Mafia Game Setting Research Using Game Refinement Measurement.'' \textit{International Conference on Advances in Computer Entertainment}. Cham: \textit{Springer International Publishing}, 2017.

\bibitem{Ri2022}
Ri, Hong, et al. ``The Dynamics of Minority versus Majority Behaviors: A Case Study of the Mafia Game." \textit{Information} 13.3 (2022): 134.



\bibitem{Tanioka2021}
Tanioka, H., and Kohri, R. ``Improving the Winning Percentage of the Werewolf Team Through Collusion Strategies." In \textit{2021 10th International Congress on Advanced Applied Informatics (IIAI-AAI)}, IEEE, July 2021, pp. 944-945.


\bibitem{Nash1950}
Nash, John F. ``Equilibrium Points in n-Person Games.'' \textit{Proceedings of the National Academy of Sciences}, vol. 36, no. 1, 1950, pp. 48-49.

\bibitem{Nash1951}
Nash, John F. ``Non-Cooperative Games.'' \textit{Annals of Mathematics}, vol. 54, no. 2, 1951, pp. 286-295.

\bibitem{Harsanyi1967}
Harsanyi, John C. ``Games with Incomplete Information Played by `Bayesian' Players, I-III Part I.'' \textit{Management Science}, vol. 14, no. 3, 1967, pp. 159-182.

\bibitem{Selten1965}
Selten, Reinhard. ``Spieltheoretische Behandlung Eines Oligopolmodells mit Nachfrageträgheit.'' \textit{Zeitschrift für die gesamte Staatswissenschaft / Journal of Institutional and Theoretical Economics}, vol. 121, no. 2, 1965, pp. 301-324.

 
\end{thebibliography}
\end{document}